\documentclass[twocolumn,superscriptaddress,aps]{revtex4-1}
\usepackage{amsmath,amssymb,bm,graphicx,color,gensymb,bbold,appendix,hyperref}

\usepackage{multirow}
\usepackage{lipsum}

\begin{document}

% ==============================================================================
\title{Rethinking $\alpha$-RuCl$_3$}
% ==============================================================================

% ==============================================================================
\author{P. A. Maksimov}
\affiliation{Bogolyubov Laboratory of Theoretical Physics, Joint Institute for Nuclear Research, 
Dubna, Moscow region 141980, Russia}
\author{A. L. Chernyshev}
\affiliation{Department of Physics and Astronomy, University of California, Irvine, California 92697, USA}
% ==============================================================================
\date{\today}
% ==============================================================================
\begin{abstract}
We argue that several empirical constraints strongly restrict parameters of the effective microscopic spin model 
describing $\alpha$-RuCl$_3$. In particular, such constraints dictate a substantial \emph{positive}  
off-diagonal anisotropic coupling, $\Gamma^\prime\!>\!0$, not anticipated previously.
The renormalization by quantum fluctuations allows to reconcile larger values of the advocated 
bare parameters with their earlier assessments and provides a consistent description 
of the field evolution of spin excitations in the paramagnetic phase. We assert that  
large anisotropic terms inevitably result in strong anharmonic coupling of magnons, 
necessarily leading to broad features in their 
spectra due to decays, in accord with the observations in $\alpha$-RuCl$_3$.
Using duality transformations, we explain the origin of the pseudo-Goldstone mode  
that is ubiquitous to the studied parameter space and is present in $\alpha$-RuCl$_3$.
Our analysis offers a description of $\alpha$-RuCl$_3$ as an easy-plane ferromagnet with
antiferromagnetic further-neighbor and strong off-diagonal couplings, which is in a fluctuating zigzag ground 
state proximate to an incommensurate phase that is continuously connected to a ferromagnetic one.
\end{abstract}
% ==============================================================================
%\pacs{75.10.Jm, 	% Quantized spin models, including quantum spin frustration
%     75.40.Gb,     % Dynamic properties
%      78.70.Nx     % Neutron inelastic scattering
%}
\maketitle
% ==============================================================================
\section{Introduction}
\label{Sec_intro}
% ==============================================================================

Finding strong physical bounds on the parameters of a model  
is often the key to understanding  the system~\cite{anthropic}. 
In quantum magnets, a nearly exact determination of their microscopic models can be achieved by   %often
measuring  the spectrum of spin excitations in magnetic fields that are 
high enough to quench quantum fluctuations \cite{RaduCsCuCl,Mourigal1D,MM1}.
Recent remarkable high-field experiments in the rare-earth pyrochlore Yb$_2$Ti$_2$O$_7$ and the subsequent 
theoretical expos\'{e} of the unfolding quantum effects in lower fields \cite{Ross11,RaduYbTiO,JeffYbTiO}
provided a spectacular demonstration of the unequivocal power of such an approach.  

However,  in anisotropic-exchange magnets, quantum effects  often remain significant even in the nominally 
spin-polarized phases \cite{RaduYbTiO,JeffYbTiO,BroholmYTO17,ColdeaEssler,Tsirlin_Review}, 
making magnetic fields necessary to  eliminate quantum fluctuations prohibitively high.
Moreover, it is common for the spin models of these materials to contain many non-negligible %symmetry-allowed 
 terms that create a multi-dimensional parameter space and make it harder to find a unique set of 
microscopic constraints \cite{Tsirlin_Review,gingrasRMP,RauReview16,WinterReview}. 
Such is the case of $\alpha$-RuCl$_3$, a honeycomb-lattice quantum magnet of great current interest 
because of its purported proximity to a spin-liquid state 
\cite{WinterReview,Plumb,Coldea15,Sandilands15,Cao16,Zvyagin17,Takagi_review,Baltz19,kaibValenti19,nagler16}.

Because of the Kitaev spin-liquid solution with much-desired topological excitations, 
the  research on $\alpha$-RuCl$_3$ has been understandably  skewed toward ignoring  realistic terms 
beyond the ``Kitaev-only'' model or adding them in a somewhat homeopathic manner with a hope for a reasonable 
phenomenology \cite{hozoi16,nagler16,motome16,motome17,K1K2}. 
On the other hand, a significant effort has also been made to establish and restrict 
physical parameters of the  realistic microscopic spin model of $\alpha$-RuCl$_3$ \cite{WinterReview}.
Without the luxury of a direct determination from the high-field spectrum measurements, studies involving
symmetry considerations, first-principles calculations, and perturbative orbital model expansions 
\cite{Chaloupka16,Kee,winter16,kee16,gong17,li17,berlijn19}, 
combined with the analysis of various experimental observations  
\cite{wen17,winter17,suga18,moore18,orenstein18,gedik19,%
kim19,okamoto19,kaib19,Winter18,Vojta1,NaglerVojta18,Rachel_18} 
have led to a  broad consensus on the \emph{minimal} microscopic model of  $\alpha$-RuCl$_3$
and to a wide range of estimates for its key parameters \cite{okamoto19}.  
It is the $K$--$J$--$\Gamma$--$\Gamma'$--$J_3$, or
generalized Kitaev-Heisenberg (KH) model,
where the  symmetry-allowed terms of the nearest-neighbor exchange matrix 
are  Kitaev, Heisenberg, and the off-diagonal  $\Gamma$ and $\Gamma'$ 
exchanges \cite{RauG,RauGp}, and $J_3$ is the 
third-neighbor Heisenberg coupling \cite{WinterReview}.
Although  minimal, this model still requires a five-dimensional parameter space 
and even a reasonable agreement on the parameter values is yet to emerge.

In this work, we use theoretical insights into several observables to strongly restrict parameters of the minimal 
model of $\alpha$-RuCl$_3$. In particular,   ESR and THz experiments on magnetic excitations in high 
fields \cite{Zvyagin17,kaib19} put a clear \emph{lower} bound on a combination
of $\Gamma$ and $\Gamma'$. Moreover,  critical fields $H^{(a)}_{c}$ and $H^{(b)}_{c}$ of the transition 
to a paramagnetic phase for the two principal in-plane directions   are nearly identical \cite{NaglerVojta18}, binding  
$\Gamma$ and $\Gamma'$ together, dictating a substantial $\Gamma'$, and also limiting  $\Gamma$-term from above. 
Similarly, the observed values of $H^{(a/b)}_{c}$  closely tie up a  combination of  $J$ and $J_3$ terms. 
Lastly, the restrictions on the spins' out-of-plane  
tilt angle \cite{kim19,Chaloupka16}, on the zigzag state being the ground state,  and on the bandwidth of the 
observed magnetic intensity \cite{Banerjee17,kaib19}, allow to put additional
bounds on the $K$, $J$, and $J_3$ terms.

Altogether, our analysis suggests a surprisingly large  
off-diagonal coupling $\Gamma^\prime\!\approx\!\Gamma/2\!>\!0$,
strong constraints on $\Gamma$ and on a combination of $J$ and $J_3$ with 
a rough estimate $|J|\!\approx\!J_3\!\approx\Gamma/2$, and the overall absolute values of all 
parameters that are generally larger than advocated previously. 
We would like to underscore that  parameters of an effective model can differ from 
the ones in the first-principles approaches. Namely, {\it ab-initio} further-neighbor terms get effectively
incorporated in the fewer  model parameters. This may allow 
to reconcile our positive $\Gamma^\prime$ term with the previous analyses \cite{Winter_footnote}.

Another reconciliation is with the smaller parameters in the prior estimates 
inferred from the experiments in the ordered zigzag phase \cite{winter17,orenstein18,gedik19,moore18}.  
They often provide a satisfactory description of the features below the field-induced transition 
to the paramagnetic phase, but  fail above it. This dichotomy can be rationalized as due to an effective  
reduction of the  bare parameters by quantum fluctuations  \cite{kaib19}, which are gradually 
lifted  by the field in the paramagnetic phase. 
For a representative set of the proposed parameters, we demonstrate  that  a 
mean-field approach to  quantum fluctuations  provides a consistent description 
of the field evolution of spin excitations in the paramagnetic phase that is in agreement with the ESR, THz, and 
Raman experiments \cite{Zvyagin17,kaib19}. This approximation is
further justified by a comparison to the exact  diagonalization results \cite{Winter18}.

A different set of quantum effects is also notable. 
As is advocated  in Refs.~\cite{winter17, kopietz20}, large off-diagonal terms in the  
anisotropic-exchange magnets necessarily precipitate strong anharmonic coupling of magnons, 
regardless of the  underlying magnetic order. 
These strong anharmonic interactions inevitably lead to large decay  rates of the 
higher-energy magnons into the lower-energy magnon continua \cite{RMP13},  
such that some of the magnon modes cease to be well-defined, leading to characteristic 
broad features in the neutron-scattering spectra. 
We apply the analysis of  Ref.~\cite{winter17} to the representative sets of our model parameters
and demonstrate a coexistence  of the low-energy well-defined quasiparticles with the 
broadened excitation continua. These results are in agreement with the prior studies \cite{winter17, kopietz20} 
and are also in accord with the experiments in $\alpha$-RuCl$_3$ \cite{nagler16,wen17,Banerjee17,banerjee18}.
Our results underscore the importance of taking into account magnon decays
in interpreting broad features in the  spectra of the strongly-anisotropic magnets \cite{JeffYbTiO}. 

There are other persistent features in the spectrum of the generalized KH model 
throughout the advocated parameter space that are also present in $\alpha$-RuCl$_3$.
One of them is the quasi-Goldstone modes 
that occur away from the ordering vector of the underlying zigzag phase \cite{wen17,banerjee18}, 
suggesting accidental near-degeneracy due to a hidden symmetry. 
We provide an insight into its nature using duality transformations of the model. 
First, a global rotation in the plane of magnetic ions transforms the generalized KH model into itself, but 
with the dominant ferromagnetic  $J\!<\!0$,  smaller  \emph{positive} and nearly equal $K$ and $\Gamma'$ terms, 
and a much smaller $\Gamma$-term.  It is important to note that this description  
is \emph{identical} to the original one and represents a feature of the $KJ\Gamma\Gamma'$ parametrization 
of  the exchange matrix.
We then show that the  Klein duality  \cite{ChKh15} transforms
the $K$--$J$--$\Gamma'$ model with $\Gamma\!=\!0$ into a $K$--$J$--$\breve{\Gamma}'$ 
model with an \emph{anti}-symmetric  $\breve{\Gamma}'$ term that is akin to the 
Dzyaloshinskyi-Moriya coupling. This last model preserves a Goldstone mode of the pure $K$--$J$ model, in a 
close similarity to the observation made for the same model on the triangular lattice \cite{Zhu19}. 

Not only does this observation explain the ubiquitous accidental pseudo-Goldstone modes, but it
also suggests a simpler model for $\alpha$-RuCl$_3$, 
which is more amendable to a detailed exploration because of the lower dimensionality of its parameter space:
the $K$--$J$--$\Gamma'$--$J_3$ model
obtained by the first transformation described above.
Moreover, the original $K$--$J$--$\Gamma$--$\Gamma'$--$J_3$ model
can be rewritten in the ``spin-ice'' language \cite{Ross11,Tsirlin_Review,Zhu19,ChKh15} that uses more
natural  spin axes tied to the honeycomb plane, yielding 
the so-called $XXZ$--${\sf J_{\pm\pm}}$--${\sf J_{z\pm}}$ form of the model.
For the parameter range that we advocate for 
$\alpha$-RuCl$_3$, the model  in this language consistently has two nearly vanishing terms, 
the $XXZ$ anisotropy  $\Delta$ and one of the anisotropic terms 
${\sf J_{\pm\pm}}$.
That is, the model that closely describes $\alpha$-RuCl$_3$ is dominated by an easy-plane 
ferromagnetic ${\sf J_1}$ and a  sizable anisotropic ${\sf J_{z\pm}}$ terms. 
Such a ${\sf J_1}$--${\sf J_{z\pm}}$--$J_3$ model description offers a  much simpler 
way of thinking about $\alpha$-RuCl$_3$,  can give a new perspective 
on its physics, and deserves further investigation.

The paper is organized as follows. We discuss the model, its parameters, their empirical constraints, 
and outline the resulting parameter space in Sec.~\ref{Sec_model}.  
In Sec.~\ref{Sec_quantum},
we discuss the effects of quantum fluctuations on  magnons in the paramagnetic and zigzag phases.
Sec.~\ref{Sec_duality} is devoted to the dual models for 
the  advocated parameter space and to different ways of representing them.
We conclude by a brief discussion in Sec.~\ref{Sec_conclusions} and provide some further details in 
Appendixes. 
\vspace{-0.2cm}

% ==============================================================================
\section{Parameters and constraints}
\label{Sec_model}
% ==============================================================================

% ==============================================================================
\begin{figure}
\centering
\includegraphics[width=0.99\linewidth]{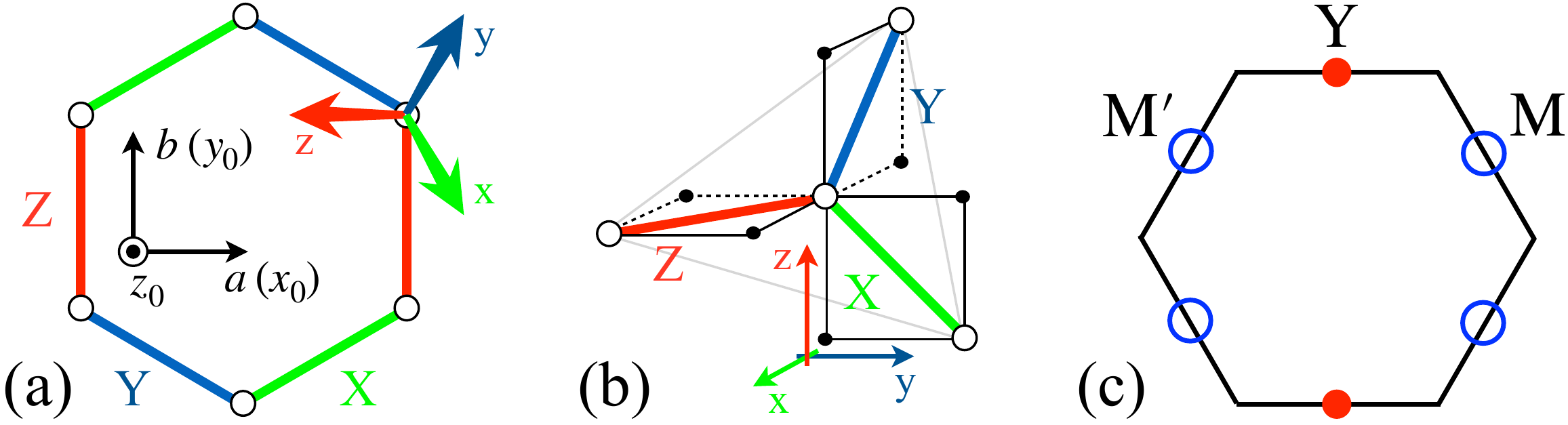}
\caption{(a) The nearest-neighbor Ru-Ru $\{{\rm X,Y,Z}\}$ bonds, crystallographic $\{x_0,y_0,z_0\}$ and cubic 
$\{\rm x,y,z\}$ axes, and principal in-plane $a(b)$ directions. (b) Cubic axes and idealized Ru-Cl bonds. 
(c) Brillouin zone with the ordering vectors of the zigzag phase Y, M, and M'.}
\label{fig_axes}
\vskip -0.2cm
\end{figure}
% ==============================================================================

The postulated minimal microscopic two-dimensional (2D) spin model of  $\alpha$-RuCl$_3$ 
is the $K$--$J$--$\Gamma$--$\Gamma'$--$J_3$ or
generalized Kitaev-Heisenberg model \cite{kee16,winter16,okamoto19},
\begin{align}
\hat{\cal H}=\hat{\cal H}_1+\hat{\cal H}_3
=\sum_{\langle ij\rangle} \mathbf{S}^{\rm T}_i \hat{\bm J}_{ij} \mathbf{S}_j
+J_3\sum_{\langle ij\rangle_3} \mathbf{S}_i \cdot \mathbf{S}_j \, ,
\label{eq_Hij}
\end{align}
where ${\bf S}_i^{\rm T}=\left(S_i^{x},S_i^{y},S_i^{z}\right)$, the third-neighbor exchange is assumed isotropic, and 
$\hat{\bm J}_{ij}$ is the nearest-neighbor bond-dependent exchange matrix. 
Since the spin-rotational symmetries in the anisotropic-exchange Hamiltonians are, generally, 
absent, the allowed matrix elements of $\hat{\bm J}_{ij}$
are  determined solely by the symmetry of the lattice \cite{RauGp}.

For $\alpha$-RuCl$_3$ and related materials \cite{WinterReview}, 
the conventional choice of the Cartesian reference frames for the  spin projections 
are the so-called cubic axes, see Fig.~\ref{fig_axes}.
They correspond to an idealized undistorted octahedral environment
of Ru$^{3+}$ and are \emph{not} coincidental with the plane of  magnetic ions, the point that is often 
lost on a non-expert or a casual reader.
These axes are natural within the orbital model considerations \cite{WinterReview},
leading  to a parametrization of the exchange matrix $\hat{\bm J}_{ij}$
that converts the nearest-neighbor part of the model (\ref{eq_Hij}) into
\begin{align}
\mathcal{H}_1=&\sum_{\langle ij \rangle_\gamma} \Big[
J \mathbf{S}_i \cdot \mathbf{S}_j +K S^\gamma_i S^\gamma_j 
+\Gamma \left( S^\alpha_i S^\beta_j +S^\beta_i S^\alpha_j\right)\nonumber
\\
\label{H_JKGGp}
&\ \ \ \ \ +\Gamma' \left( S^\gamma_i S^\alpha_j+S^\gamma_i S^\beta_j+S^\alpha_i S^\gamma_j
+S^\beta_i S^\gamma_j\right)\Big],
\end{align}
where  $\langle ij \rangle_\gamma$ numerates the bonds $\gamma=\{{\rm X,Y,Z}\}$, 
with the triads of $\{\alpha,\beta,\gamma\}$ being $\{{\rm y,z,x}\}$ on the X bond, $\{{\rm z,x,y}\}$ on the Y bond, 
and $\{{\rm x,y,z}\}$ on the Z bond, respectively, see Fig.~\ref{fig_axes} for the cubic axes, crystallographic reference
frame $\{x_0,y_0,z_0\}$, and other notations. We also note that the 
parametrization of the exchange matrix $\hat{\bm J}_{ij}$ that is used in (\ref{H_JKGGp})
is a subject of some less-than-obvious transformations \cite{ChKh15} under relatively trivial symmetry 
operations discussed in Sec.~\ref{Sec_duality}. 

A number of the $\{J,K,\Gamma,\Gamma',J_3\}$ parameter sets have been proposed to describe $\alpha$-RuCl$_3$  
using the first-principles methods  \cite{winter16,kee16,gong17,li17,berlijn19,hozoi16} 
and phenomenological analyses 
\cite{wen17,winter17,suga18,moore18,orenstein18,gedik19,kim19,okamoto19,kaib19,nagler16}.
We provide a  compilation of them in Table~\ref{table1} in the end of this Section and compare with the 
ranges  advocated in the present study.
The coupling that is believed to be the leading one is the (negative) Kitaev 
term, $K\!<\!0$. The off-diagonal $\Gamma\!>\!0$ term is also  discussed as 
significant and potentially comparable to $|K|$, while the ferromagnetic exchange  $J\!<\!0$ is believed to be 
subleading \cite{WinterReview}. 
All three ``main'' parameters  vary quite significantly between the studies, with  
the antiferromagnetic third-neighbor $J_3$ of the same order as $|J|$ also frequently invoked, 
and a small, predominantly negative $\Gamma'$  included as being allowed by symmetry 
\cite{RauGp,winter16,okamoto19}. 

In the following, we use the first-principle guidance for $\alpha$-RuCl$_3$ \cite{winter16} and 
assume that $K\!<\!0$. We  also use other restrictions from these works, such as some of the prevalent 
hierarchies of the couplings. However, we demonstrate that it is the currently available 
phenomenology that is powerful enough to significantly restrict and drastically revise the  physically  reasonable 
parameter space of the generalized KH model for $\alpha$-RuCl$_3$.

% ==============================================================================
\subsection{ESR and THz data}
\label{Sec_constraints}
% ==============================================================================

The electron spin resonance (ESR),  terahertz  (THz), and Raman  spectroscopies 
have provided detailed information on the 
$\mathbf{q}\!=\!0$ magnetic excitations of $\alpha$-RuCl$_3$ and their field evolution  
 in the fluctuating paramagnetic state   \cite{Zvyagin17,kaib19,Wulferding19}. 
While a rich spectrum with multiple modes has been analyzed \cite{kaib19}, we focus on the field-dependence of  
the low-energy single-magnon mode \cite{F_footnote}. 

\begin{figure}
\centering
\includegraphics[width=0.99\linewidth]{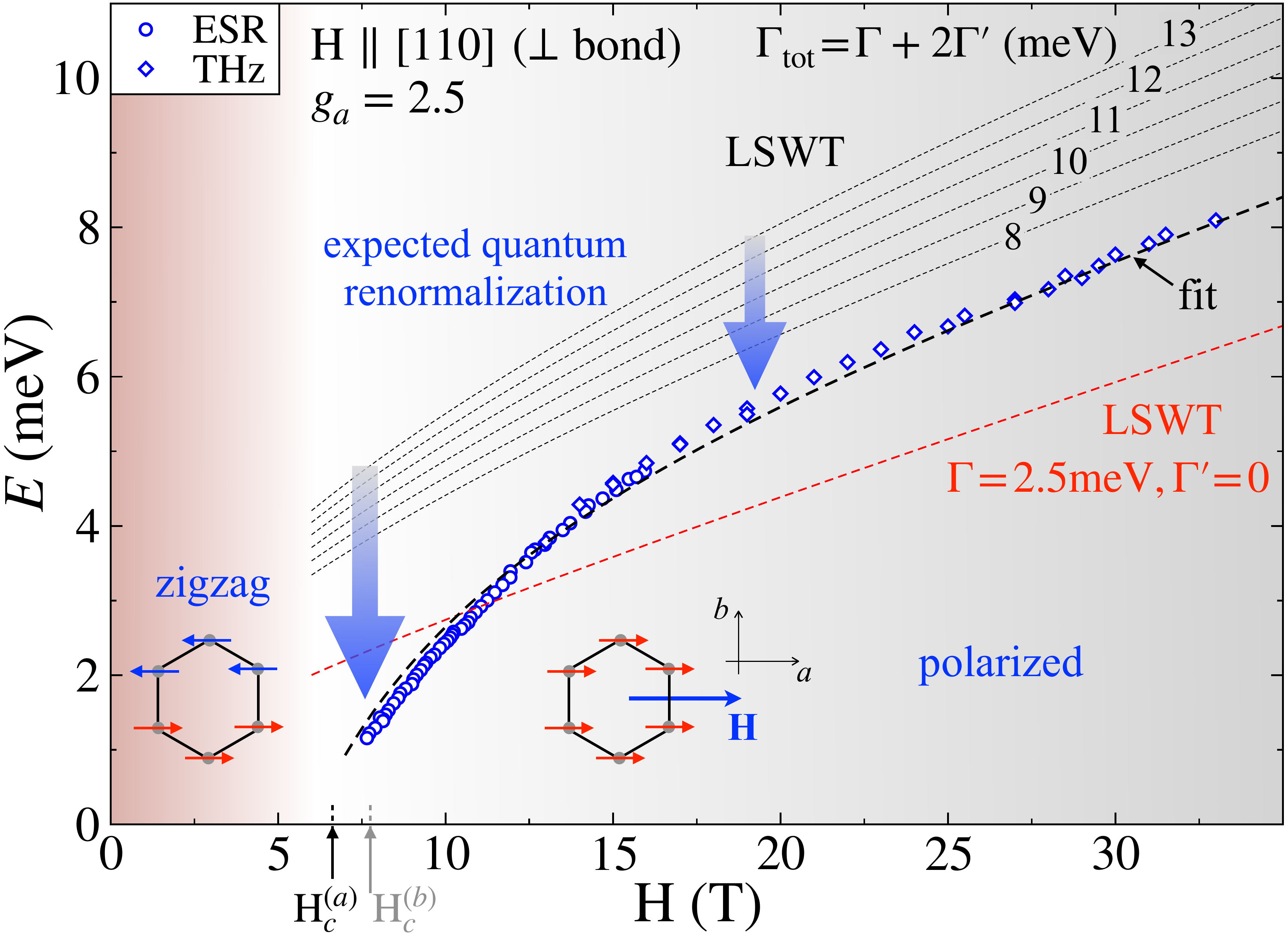}
\vskip -0.2cm
\caption{ESR \cite{Zvyagin17} and THz \cite{kaib19} data and their fit for the magnon energy gap, $\varepsilon_0$, 
at $\mathbf{q}\!=\!{\bf 0}$ vs field in the $a$-direction; LSWT results from Eq.~(\ref{Ek0}) 
for representative $\Gamma_{\rm tot}$ from 8 meV to 13 meV and for $\Gamma\!=\!2.5$ meV \cite{winter17} for
a comparison.  Arrows indicate anticipated downward renormalization of the LSWT results by quantum fluctuations.
Insets: sketches of the zigzag and polarized states and  in-plane $a$ and $b$ directions.}
\label{fig_ESRfit}
\vskip -0.2cm
\end{figure}

In Fig.~\ref{fig_ESRfit}, we show the data for  this mode from the ESR (Ref.~\cite{Zvyagin17}) 
and THz (Ref.~\cite{kaib19}) studies 
for the in-plane field direction that is  perpendicular to the Ru-Ru bond, referred to as the $a$-direction,
for the field range from the critical field $H^{(a)}_{c}\!\approx\! 6\text{T}$  to 35T.
The data for the field along the $b$-direction are quite similar, suggesting nearly equal 
$g$-factors, the point also supported by the earlier  studies \cite{hozoi16,Winter18}.
We provide a fit of the data by 
\begin{align}
\label{E0_exp}
\varepsilon_0\!=\!h + a_0 + a_1/h\,,
\end{align}
with $h\!=\!g\mu_BH$, $a_0\!=\!4.2$ meV, and $a_1\!=\!30$ meV$^2$, 
which is motivated by the high-field expansion of Eq.~(\ref{Ek0}) below.  
Throughout this work, we use $g_a\!=\!g_b\!=\!2.5$, which is in accord with the previous 
 estimates \cite{hozoi16,Winter18,Kubota15}.

% ==============================================================================
% ==============================================================================

Importantly, the linear spin-wave theory (LSWT) gives the $\mathbf{q}\!=\!0$ magnon energy   that
 depends  on a combination of only two parameters of the model (\ref{eq_Hij}), 
$\Gamma_{\rm tot}\!=\!\Gamma+2\Gamma'$,
\begin{align}
\label{Ek0}
\varepsilon^{(0)}_{0}=\sqrt{h \big(h+3S \big(\Gamma+2\Gamma'\big)\big)},
\end{align}
where $h\!=\!g\mu_BH$. This result  is asymptotically exact in the $H\!\rightarrow\!\infty$ limit 
where fluctuations are suppressed. Expansion of (\ref{Ek0}) in $1/h$ yields the form used in (\ref{E0_exp}).

In Fig.~\ref{fig_ESRfit}, we present LSWT results  for several representative  $\Gamma_{\rm tot}$ from 8 to 13 meV. 
The lowest LSWT line ($\Gamma\!=\!2.5$ meV) uses parameters that were 
successful in describing the low-field phenomenology of  $\alpha$-RuCl$_3$ \cite{winter17,moore18}, but clearly 
fail to reproduce the high-field data, suggesting significantly larger   $\Gamma_{\rm tot}$.
The  key point is that one should generally expect a \emph{downward} renormalization of the LSWT spectrum 
due to quantum fluctuations \cite{winter17,tri09}, 
as is confirmed by a comparison with the  exact  diagonalization results of Ref.~\cite{Winter18} in 
Sec.~\ref{Sec_quantum}.  

Therefore, it is clear from Fig.~\ref{fig_ESRfit}  that $\Gamma_{\rm tot}$ cannot be less than $\approx\!8$ meV
and it is also hard to justify it to be larger than $\approx\!13$ meV as this would imply unphysically large 
fluctuations in a strongly gapped high-field state. 
Thus, while the latter is not a precise constraint, 
there is a clear sense of both the lower and the upper bounds 
on the   value of $\Gamma+2\Gamma'$ from the ESR and THz data.

Qualitatively, fluctuations produce the downward shift of the spectrum due to  repulsion of the one- and two-magnon 
states, which is expected to get stronger near the critical field, 
in agreement with Fig.~\ref{fig_ESRfit} and with a discussion in Ref.~\cite{kaib19}.
We also note, that  the observed single-magnon energy can be  related to the ``bare'' LSWT result 
of Eq.~(\ref{Ek0}) as $\varepsilon_0\!=\!\Lambda\varepsilon^{(0)}_{0}$, where  $\Lambda$ is the field-dependent
renormalization factor with the high-field behavior $\Lambda\!=\!1-O(h^{-1})$.
Thus, while naively one can extract $\Gamma_{\rm tot}$  using expansion in Eq.~(\ref{E0_exp}) directly
from the $a_0$ term, the fluctuation factor provides a significant correction to it that requires a self-consistent 
consideration.

As we show in Sec.~\ref{Sec_quantum}, the field-dependent
renormalization factor can be approximated by the reduced  ordered moment, 
$\Lambda\!=\!\langle S\rangle/S$, as follows from the self-consistent random-phase approximation (RPA) \cite{Tyablikov}.
According to it,  fluctuation corrections to $\varepsilon^{(0)}_{0}$ at higher fields can still produce a  substantial 
downward shift, suggesting  the lower limit for $\Gamma_{\rm tot}$ to be $\agt 9$ meV. 

\vspace{-0.3cm}

% ==============================================================================
\subsection{Critical fields}
% ==============================================================================
\vskip -0.2cm

In $\alpha$-RuCl$_3$, the in-plane field induces a phase transition from the zigzag to a fluctuating paramagnetic state
at a critical field about $7$T \cite{Winter18,Cao16,NaglerVojta18}, see Fig.~\ref{fig_ESRfit}. 
An additional transition at a lower field \cite{NaglerVojta18} has been identified with an 
interplane ordering \cite{Baltz19,Vojta3D} and is unrelated to the key physics of $\alpha$-RuCl$_3$ discussed in this 
work \cite{footnote3D}. 
 
In the field-induced paramagnetic phase,   magnon spectrum is gapped and the  transition to 
the  zigzag phase upon lowering  the field corresponds to a softening of the spectrum. 
The gap closes at the ordering vectors associated with the zigzag structure,
the face-centers of the Brillouin zone, see Fig.~\ref{fig_axes}(c). 
For $H\!\parallel\! b$, the ordering vector of the single field-selected zigzag domain is Y, and
for $H\!\parallel\! a$, the two domains have the ordering vectors at the M and M$'$ points, respectively 
\cite{Sears17,orenstein18,Vojta3D}.

In the  paramagnetic phase spins are oriented along the field and the magnon spectrum 
can be obtained analytically, see Appendix \ref{app_A}. 
The condition on the gap closing yields the critical fields for $H\!\parallel\! a$  and $H\!\parallel\! b$, 
see also \cite{footnote_Hc}
\begin{align}
\label{Hca}
h_c^{(a)}&=J+3J_3+\frac{1}{12}\big(5K-5\Gamma-16\Gamma'\big)\\
&+\frac{1}{12}\sqrt{\big(K+5\Gamma+4\Gamma'\big)^2+24 \big( K-\Gamma+\Gamma'\big)^2},\nonumber\\
\label{Hcb}
h_c^{(b)}&=J+3J_3+\frac{1}{4}\big(2K-\Gamma-6\Gamma'\big)\\
&+\frac{1}{12}\sqrt{\big(2K+7\Gamma+2\Gamma'\big)^2+32 \big( K-\Gamma+\Gamma'\big)^2}, \nonumber
\end{align}
where $h_c^{(\alpha)}\!=\!g_\alpha \mu_B H_c^{(\alpha)}$. An important feature of these results is that the 
\textit{difference} of the critical fields in (\ref{Hca}) and (\ref{Hcb}) appears to be a function 
of only three anisotropic terms of the model: $K,\Gamma$, and $\Gamma'$. 
As is discussed above, we assume the $g$-factors in the two principal directions to be the same, so 
$\Delta h_c\!=\!g\Delta H_c$, with $\Delta H_c\!=\!H_c^{(b)}\!-\!H_c^{(a)}$. 
This feature is key to the constraints proposed below.

% ==============================================================================
\begin{figure}
\centering
\includegraphics[width=0.99\linewidth]{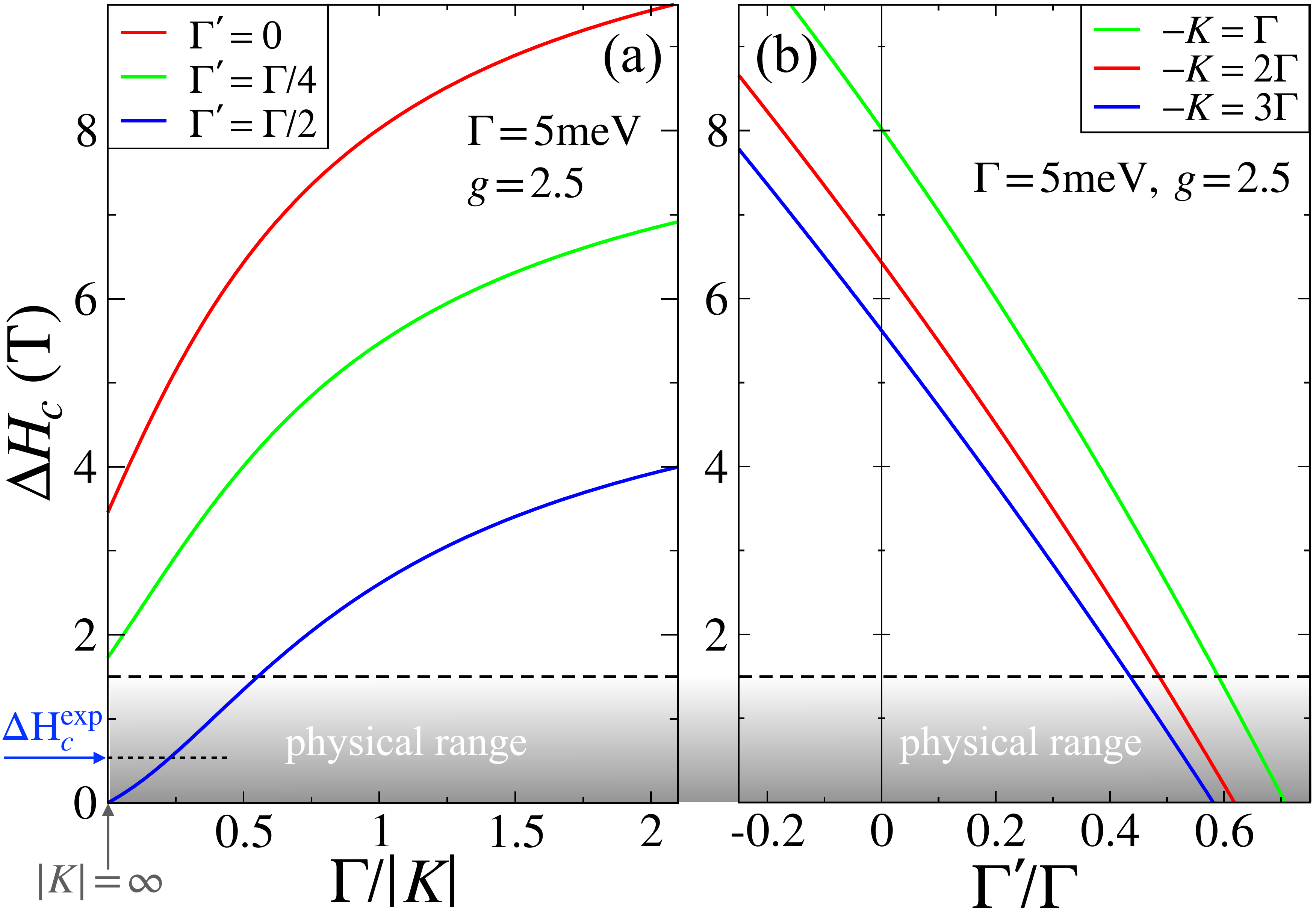}
\vskip -0.2cm
\caption{(a) The difference of the critical fields  
from Eqs.~(\ref{Hca}) and (\ref{Hcb}), $\Delta H_c\!=\!H_c^{(b)}\!-\!H_c^{(a)}$,  vs $\Gamma/|K|$ 
for representative values of $\Gamma'/\Gamma$ and  $\Gamma\!=\!5$ meV. 
(b) $\Delta H_c$  vs $\Gamma'/\Gamma$ for  representative ratios of $K/\Gamma$.
The  physical range for $\Delta H_c$ is $<\!1.5$T, see text; the experimental value is 
$\Delta H_c^{\rm exp}\!=\!0.6$T \cite{NaglerVojta18}.
The physical range requires significant $\Gamma'\!\sim\!\Gamma/2\!>\!0$.}
\label{fig_dHvsK}
\vskip -0.4cm
\end{figure}
% ==============================================================================

Before we discuss them in more detail, we note that, 
experimentally, the critical fields in $\alpha$-RuCl$_3$ for $a$- and $b$-directions  are nearly identical 
\cite{Zvyagin17,NaglerVojta18}. 
While small   $\Delta H_c$ seems to be a minor point, it is virtually impossible
to reproduce from  Eqs.~(\ref{Hca}) and (\ref{Hcb})  without a sizable $\Gamma'$, 
the difficulty also clearly encountered in Ref.~\cite{NaglerVojta18} that used a  model with $\Gamma'\!=\!0$. 

We demonstrate this by Fig.~\ref{fig_dHvsK}(a) showing $\Delta H_c$ vs $\Gamma/|K|$ 
for several values of $\Gamma'/\Gamma$ and for a representative value of $\Gamma\!=\!5$ meV, see Table~\ref{table1}. 
The highlighted physical range for $\Delta H_c$ is chosen as $<\!1.5$T to account for possible difference 
of the $g$-factors, while the experimental value is $\Delta H_c^{\rm exp}\!=\!0.6$T \cite{NaglerVojta18}.
It is clear from Fig.~\ref{fig_dHvsK}, that even at $|K|\!\rightarrow\! \infty$ the asymptotic value of $\Delta H_c$
for $\Gamma'\!=\!0$ is well above the physical range and  a positive $\Gamma'\!\agt\!\Gamma/2$ is needed to reach it. 
The same effect is illustrated in Fig.~\ref{fig_dHvsK}(b), with $\Delta H_c$ plotted vs $\Gamma'/\Gamma$ 
for three representative ratios of $K/\Gamma$. Again, a model without a significant positive $\Gamma'$ 
cannot reproduce observed small difference between the critical fields. 

Superficially, a large positive $\Gamma'$  contradicts first-principles results for  $\alpha$-RuCl$_3$ 
\cite{winter16,kee16,berlijn19}. However, as we discussed in Sec.~\ref{Sec_intro}, 
our model implicitly incorporates further-neighbor terms into fewer effective parameters, 
with a phenomenology dictating physical answer. This result is also in accord with
the ESR/THz constraints that require large $\Gamma\!+\!2\Gamma'$. Having
substantial $\Gamma'\!\sim\!\Gamma/2$ removes the need for the unphysically large $\Gamma$ 
in explaining some of the other $\alpha$-RuCl$_3$ phenomenologies \cite{wen17,kim19,okamoto19}.

One concern is the potential effect of quantum fluctuation corrections on the LSWT
results for the critical fields in Eqs.~(\ref{Hca}) and (\ref{Hcb}). 
However, such corrections are  unlikely to affect the smallness of their difference, 
$\Delta H_c\!\ll\!H_c^{(a/b)}$,
and the arguments on a sizable $\Gamma'$ that follow from it. 
Moreover, the self-consistent mean-field RPA approach advocated in Sec.~\ref{Sec_quantum}
predicts no quantum effects on the critical fields. Near the transition, Zeeman energy of the fluctuating spin 
polarization, $(H-H_c)\langle S\rangle$, competes with the gap that is  reduced by quantum fluctuations,
$\Lambda\Delta^{(0)}$, where $\Lambda\!=\!\langle S\rangle/S$ as suggested above \cite{Tyablikov}. Then, 
the  condition on closing of the gap is the same as within the LSWT, in which  bare Zeeman energy, 
$(H-H_c)S$, 
competes with the bare gap, $\Delta^{(0)}$. Thus, the critical field is unchanged by the fluctuations. 
While this is a mean-field argument, it points to suppressed quantum effects on the critical fields. 

\vspace{-0.3cm}

% ==============================================================================
\subsection{Empirical constraints, I}
% ==============================================================================
\vskip -0.2cm

As is discussed above, for the model (\ref{eq_Hij}) of $\alpha$-RuCl$_3$ 
there are bounds on $\Gamma_{\rm tot}\!=\!\Gamma+2\Gamma'$
and on $\Delta H_c$. 
Moreover, $\Delta H_c$  depends only on three parameters of the  model: $\Delta H_c(K,\Gamma,\Gamma')$. 
Thus, if one would be able to fix exactly both $\Gamma_{\rm tot}$ and $\Delta H_c$, this would restrict
the 3D  parameter subspace  of $\{K,\Gamma,\Gamma'\}$  to a  1D curve.  

\begin{figure}
\centering
\includegraphics[width=0.99\linewidth]{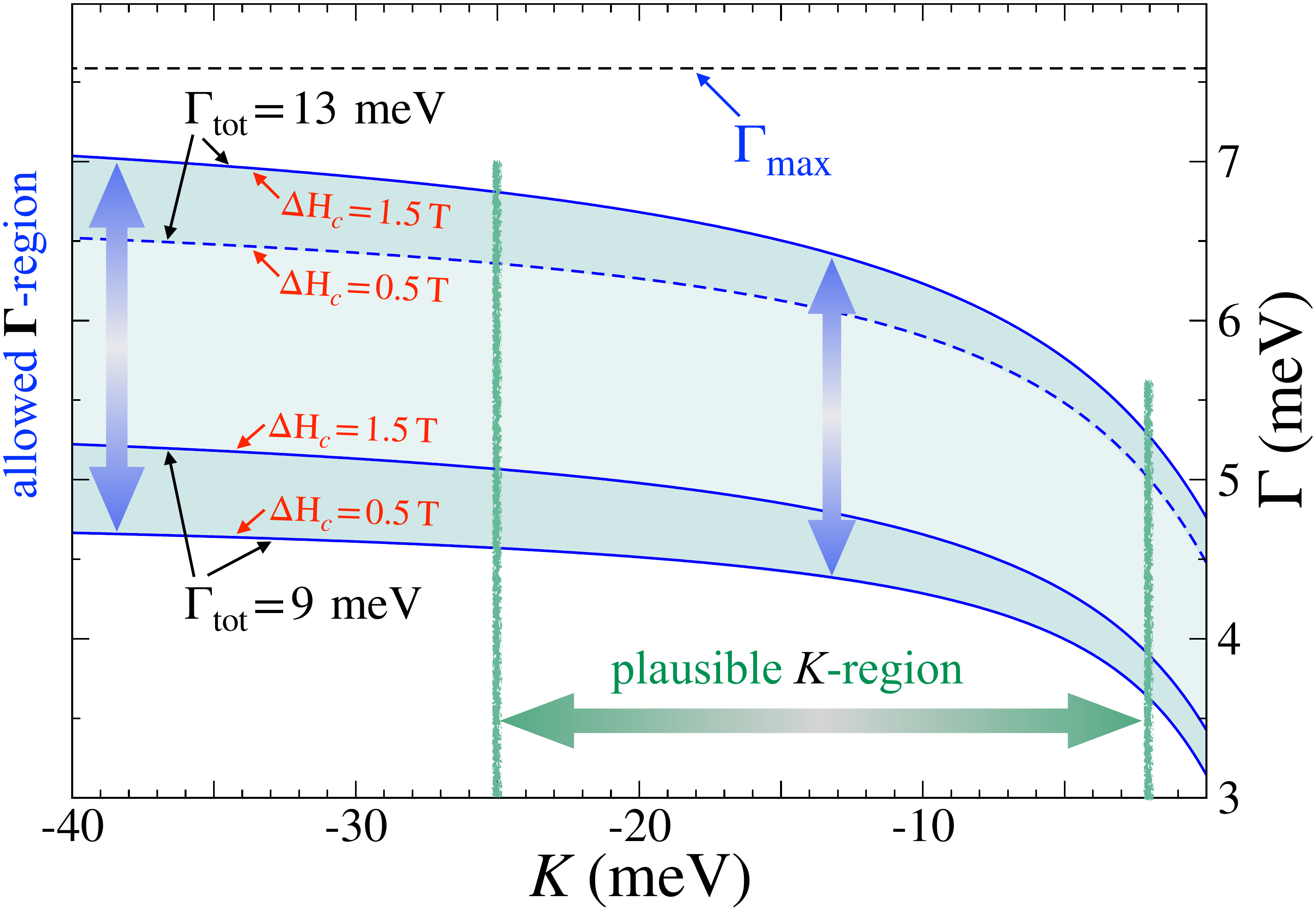}
\vskip -0.2cm
\caption{The $K$--$\Gamma$ projection of the parameter range restricted by two constraints: 
9 meV$<\!\Gamma+2\Gamma'\!<\!13$ meV and 0.5T$<\!\Delta H_c\!<\!1.5$T  (shaded area). 
$\Gamma_{\rm max}$ (dashed line) is the asymptotic value of the upper boundary of the shaded region,
$\{\Gamma_{\rm tot},\Delta H_c\}\!=\!\{13\,{\rm meV}, 1.5\,{\rm T}\}$,  
for  $|K|\!\rightarrow\!\infty$. Plausible region for $K$ suggested by the 
prior estimates, Table~\ref{table1}, is indicated.}
\label{fig_GK}
\vskip -0.4cm
\end{figure}

To get an insight into the resulting constraints, we show projections of such curves onto the $K$--$\Gamma$ plane 
 in Figure~\ref{fig_GK} for four sets of $\{\Gamma_{\rm tot},\Delta H_c\}$ 
 with  $\Gamma_{\rm tot}\!=\!9$~meV and 13 meV and $\Delta H_c\!=\!0.5$T and 1.5T. 
Obviously, the entire range of  $\Gamma_{\rm tot}$ from 9~meV to 13~meV and of $\Delta H_c$ 
from 0.5T to 1.5T are 
confined between these curves, shown by the shaded area.  
This range of $\Gamma_{\rm tot}$ is bounded by the ESR/THz as discussed above. 
Instead of fixing $\Delta H_c$ to its experimental value of 0.6T \cite{NaglerVojta18}, 
we allow for an additional range from 0.5T to 1.5T to account for small differences in the $g$-factors
and for the residual quantum corrections to $H_c^{(a)}$  and $H_c^{(b)}$in Eqs.~(\ref{Hca}) and (\ref{Hcb}). 

It is clear from Fig.~\ref{fig_GK}, 
that the values of $\Gamma$ are strongly constrained already at this stage, while there is no upper limit on $|K|$. 
An expression for  $K(\Gamma, \Gamma_{\rm tot}, \Delta H_c)$ vs $\Gamma$
indeed displays an unbounded asymptotic form $K\!\sim\!1/(\Gamma-\Gamma_{\rm max})$ with 
$\Gamma_{\rm max}\!=\!\Gamma_{\rm tot}/2+5\Delta h_c$. Figure~\ref{fig_GK} shows such $\Gamma_{\rm max}$
 by the dashed line for the upper-boundary values of  $\Gamma_{\rm tot}\!=\!13$~meV and $\Delta H_c\!=\!1.5$T.

The main message of Fig.~\ref{fig_GK} is that $\Gamma$ for the model (\ref{eq_Hij}) of $\alpha$-RuCl$_3$ 
is constrained  from both below and above mainly by the bounds on $\Gamma_{\rm tot}$ from the ESR/THz gap
and to a lesser extent by the variation of allowed $\Delta H_c$, 
while $K$ is only restricted by the choice of $K\!<\!0$. However,   Kitaev term  
also has physical constraints, see Table~\ref{table1}, as we highlight in Fig.~\ref{fig_GK},  which should lead to even tighter 
bounds on the possible ranges of $\Gamma$. Overall, the ``typical'' value of $\Gamma$ appears to be 
$\sim\!\Gamma_{\rm tot}/2$, and $\Gamma'\!\sim\!\Gamma/2$.
 
% ==============================================================================
% ==============================================================================

% ==============================================================================
\begin{figure*}
\centering
\includegraphics[width=0.99\linewidth]{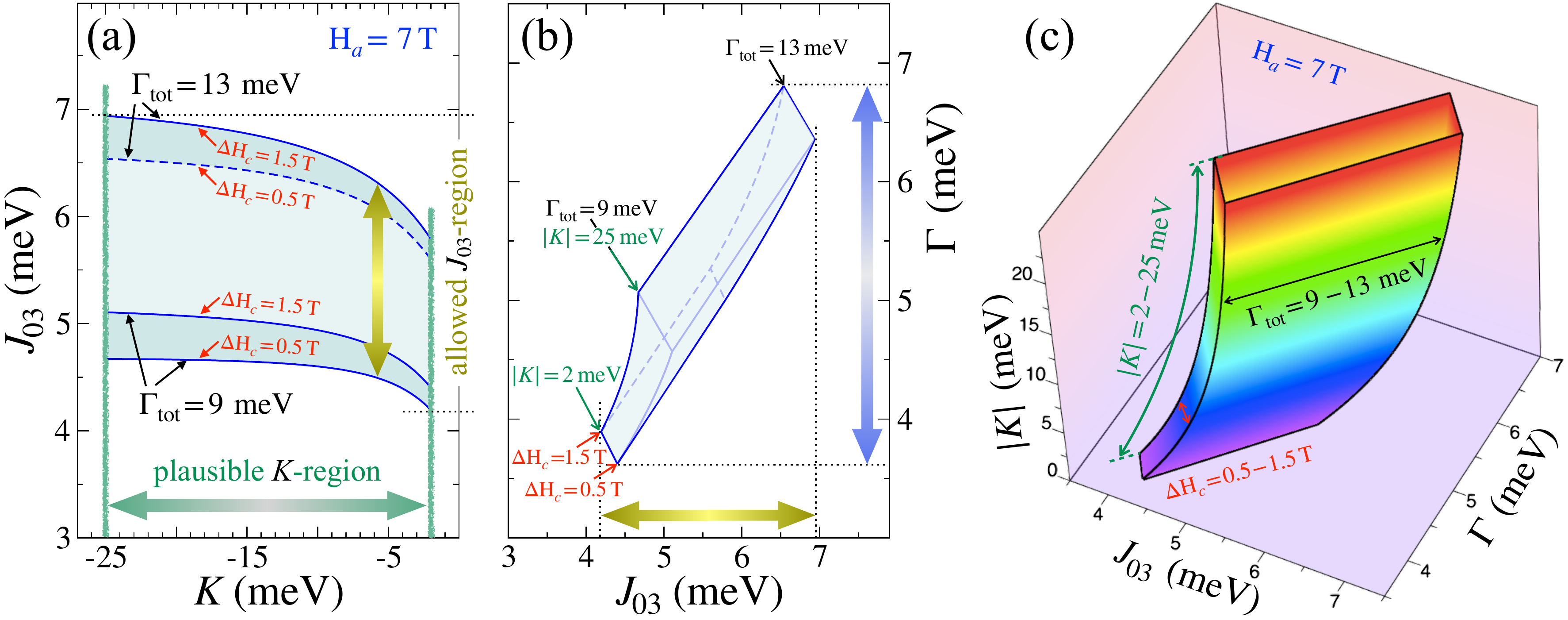}
\vskip -0.2cm
\caption{(a) The $K$--$J_{03}$ and (b) the $J_{03}$--$\Gamma$  projections of the 
parameter space allowed by  three constraints:  $9\!<\!\Gamma_{\rm tot}\!<\!13$ meV,  
$0.5\!<\!\Delta H_c\!<\!1.5$T, and $H_c^{(a)}\!=\!7$T. These criteria strongly bind $\Gamma$ and $J_{03}$, see (b). 
The $K$--$J_{03}$ projection in (a) is similar to the $K$--$\Gamma$ projection in Fig.~\ref{fig_GK}, 
with the upper and lower limits on $J_{03}$ and plausible boundaries on the Kitaev term $K$ indicated. 
(c) The 3D  $J_{03}$--$\Gamma$--$K$ subspace of the allowed parameters, with the extent of 
$\Gamma_{\rm tot}$ giving the length, $\Delta H_c$ determining the width, and $K$ restricting the height of the 
3D region. Projections of this 3D object yield Figs.~\ref{fig_GK}, \ref{fig_j03}(a) and \ref{fig_j03}(b).}
\label{fig_j03}
\vskip -0.4cm
\end{figure*}
% ==============================================================================

There is an important feature of the critical fields $H_c^{(a)}$ and $H_c^{(b)}$ given by 
Eqs.~(\ref{Hca}) and (\ref{Hcb}).
They both depend on the Heisenberg exchanges only via  a  linear combination 
\begin{align}
\label{J03}
J_{03}\equiv J+3J_3,
\end{align}
which makes $J_{03}$ a natural variable in the discussion of the empirical constraints.  
Quantitatively, this dependence is also very strong. 
Using $g\!=\!2.5$, a relatively small change of $J_3$ by 0.3 meV modifies $H_c$ by about 6T. 
Incidentally, this also mitigates concerns about quantum effects on experimental values of the critical fields compared
to the theoretical ones, as a small adjustment of $J_3$ in the latter is sufficient to match the former.

Thus, for the purpose of considering critical fields, the 5D parameter space of 
the model (\ref{eq_Hij}) of $\alpha$-RuCl$_3$ 
is effectively reduced to a 4D subspace by using $J_{03}$ from Eq.~(\ref{J03}): 
$H_c^{(\alpha)}\!=\!H_c^{(\alpha)}(K,\Gamma,\Gamma',J_{03})$.
Using the same bounds on $\Gamma_{\rm tot}$ and $\Delta H_c$ as in Fig.~\ref{fig_GK} 
and   experimental value of  $H_{c,{\rm exp}}^{(a)}\!=\!7$T \cite{NaglerVojta18}
allows us to put constraints on $J_{03}$ in a  similar manner to that of the  $\Gamma$-term. 

As one can see from Figure~\ref{fig_j03}(a),  the $K$--$J_{03}$ projection 
of the  parameter range restricted by these constraints 
shows the same characteristics as the 
 $K$--$\Gamma$ projection in Fig.~\ref{fig_GK}. That is, $J_{03}$ 
is constrained  from  below and above, while $K$ is only semi-bounded. 
In addition, the ``plausible range'' of $2\!<\!|K|\!<\!25$~{\rm meV}, see Table~\ref{table1}, 
further restricts $J_{03}$ within the ``typical'' values that are very similar to that of $\Gamma$,  
$J_{03}\!\sim\!\Gamma_{\rm tot}/2$.
We  note that these results are rather insensitive to the variations of 
$H_{c,{\rm exp}}^{(a)}$, easily so within the limits of $\pm$2T, as they can be  effectively absorbed 
into small changes of $J_{03}$ of order $\sim\!0.1$~meV.

Altogether,  constraints on the parameters of the model (\ref{eq_Hij}) of $\alpha$-RuCl$_3$ discussed so far
have resulted in strong bounds on $\Gamma$ and $J_{03}$. This is demonstrated explicitly in Figure~\ref{fig_j03}(b),
which shows a $J_{03}$--$\Gamma$ projection of the allowed parameter ranges that are restricted
by the same limits as in Figs.~\ref{fig_GK} and \ref{fig_j03}(a),  
dictated by the ESR/THz gap, variation of  $\Delta H_c$, and fixed $H_{c,{\rm exp}}^{(a)}\!=\!7$T.
In this Figure, the narrow width of the projection is mostly controlled by $\Delta H_c$, 
while the length is due to the limits on $\Gamma_{\rm tot}$ and $K$. In Fig.~\ref{fig_j03}(b), we
explicitly limited  $K$ to the ``plausible range'' of 2~meV$<\!|K|\!<$25~meV.

Lastly, all three projections of Figs.~\ref{fig_GK}, \ref{fig_j03}(a), and \ref{fig_j03}(b) are summarized as a 
3D shape in Figure~\ref{fig_j03}(c), which makes it explicit that    
the allowed regions illustrated in each  Figure correspond to a projection of this three-dimensional object 
onto a respective plane. The aforementioned correlations between different parameters also become  clearer, 
with the upper and lower bounds on $\Gamma_{\rm tot}$ providing the ranges for $\Gamma$ and  $J_{03}$, while 
the difference of the critical fields $\Delta H_c$ is giving a  narrow width of the allowed 3D parameter space. 
However,  Kitaev term remains unbounded and so do the Heisenberg exchanges 
$J$ and $J_3$, as we have only restricted their combination. 
Therefore, more empirical constraints are needed.

\vspace{-0.3cm}

% ==============================================================================
\subsection{Empirical constraints, II}
% ==============================================================================
\vskip -0.2cm

To establish further constraints for the model (\ref{eq_Hij}) of $\alpha$-RuCl$_3$, 
we employ two additional ``soft" criteria motivated by several experimental results. 
Below we discuss the range of the out-of-plane tilt angle of  spins in the zigzag state and an overall upper  
limit on the energy bandwidth of the observed magnetic intensity. 

\begin{figure}[b]
\centering
\includegraphics[width=0.99\linewidth]{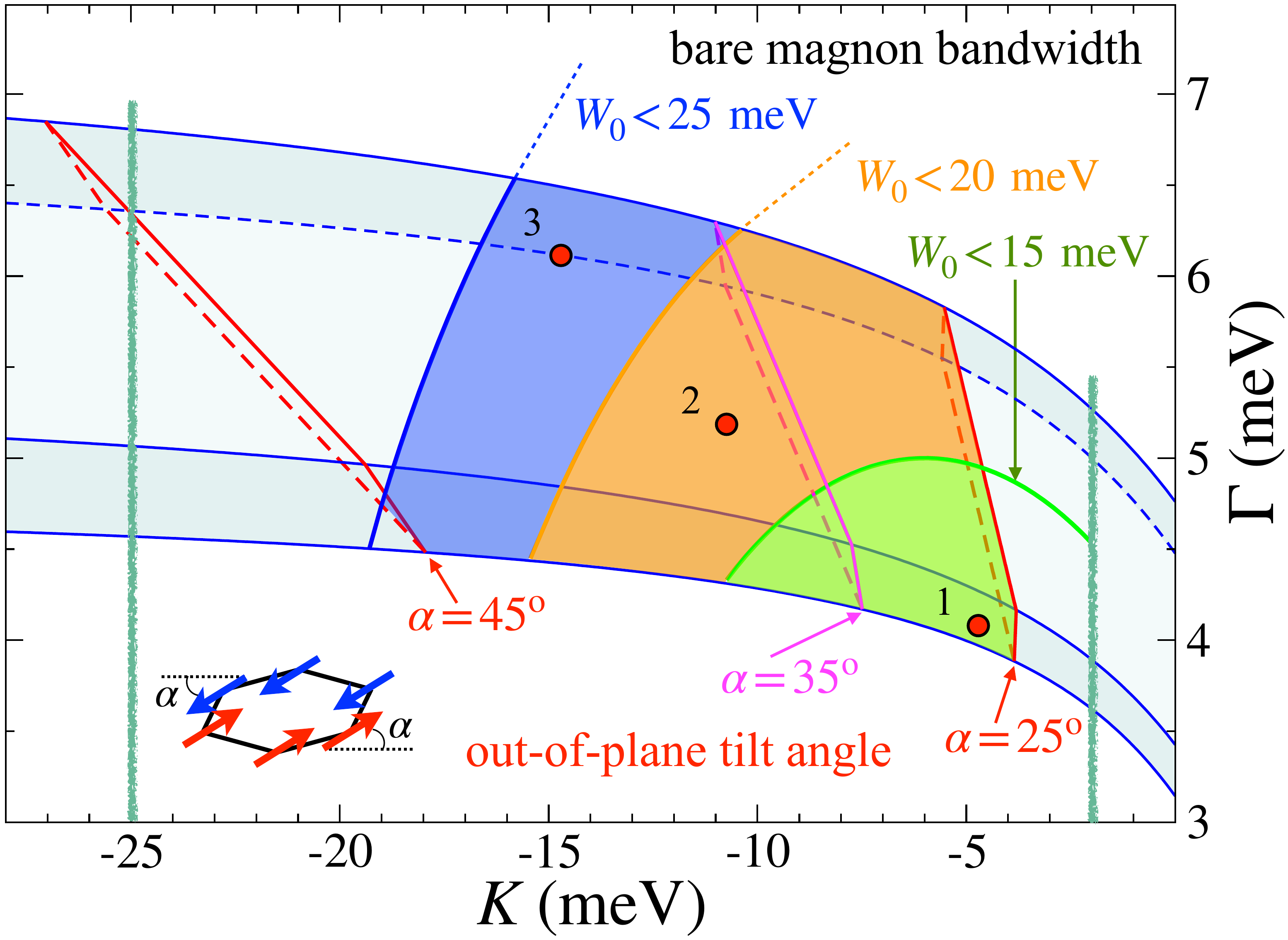}
\vskip -0.2cm
\caption{Same as Fig.~\ref{fig_GK} with constraints on  the tilt angle $\alpha$ (shown in the inset)
 and on magnetic intensity width $W_0$, see text. Dots are representative parameter sets from Sec.~\ref{Sec_model}~E.}
\label{fig_6}
\end{figure}

First of the ``soft'' criteria is the experimentally observed tilt of the spins away from the $ab$-plane 
in  zero-field zigzag ground state of $\alpha$-RuCl$_3$, see inset in Fig.~\ref{fig_6}. 
This effect has  been  discussed in Refs.~\cite{ChKh15,Chaloupka16}, 
with the tilt occurring   due to anisotropic  terms and the angle in the classical limit  given by  
\begin{align}
\tan 2\alpha= 4\sqrt{2}\, \frac{1+r}{7r-2}, \ \  \ r=-\frac{\Gamma}{K+\Gamma'}.
\label{eq_alpha}
\end{align}
It has also been analyzed by the neutron diffraction, muon spin relaxation, and  
resonant elastic x-ray scattering \cite{Cao16,Tanaka18,kim19},
with the best fits giving the tilt angle around $\alpha\!\approx\!35^\circ$, see Fig.~\ref{fig_6}.  
By comparing to exact diagonalization, it was shown in Ref.~\cite{Chaloupka16} that  quantum corrections
can modify the classical value of $\alpha$   by about 5$^\circ$. Furthermore, there may be a 
difference between the calculated direction of the pseudospin in (\ref{eq_alpha}) and the 
experimentally measured direction of the magnetic moment, 
see Ref.~\cite{Chaloupka16}. 
To account for these effects, we take a generous range of  $25^\circ\!<\!\alpha\!<\!45^\circ$ as our  criterion
instead of fixing the tilt angle to a particular value, see also Appendix~\ref{app_0}. 

% ==============================================================================
% ==============================================================================

% ==============================================================================
\begin{figure*}
\centering
\includegraphics[width=0.99\linewidth]{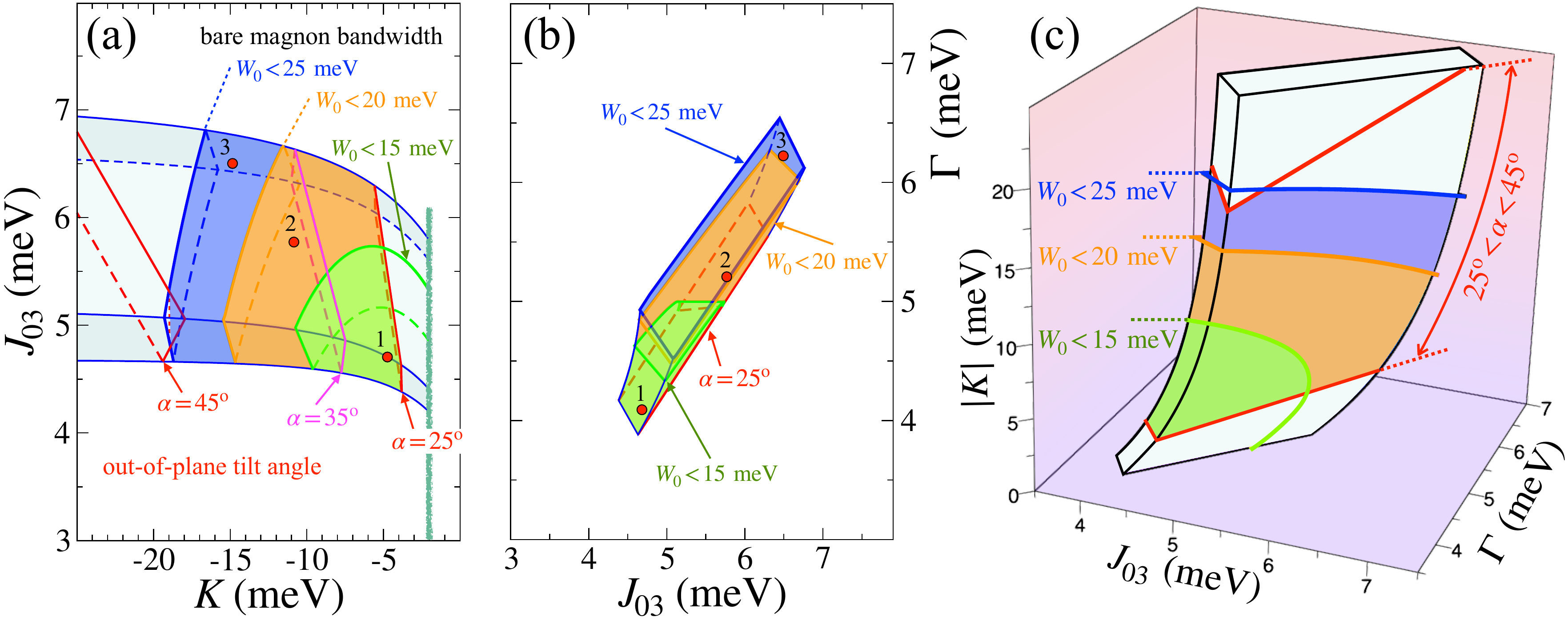}
\vskip -0.2cm
\caption{Same as Fig.~\ref{fig_j03} with  constraints on  the tilt angle $\alpha$, magnetic intensity width $W_0$ 
(``realistic'' $\alt\!15$~meV, ``generous''  $\alt\!20$~meV, and ``outrageous'' $\alt\!25$~meV), 
and zigzag ground state, see text. Dots are  parameter sets from Sec.~\ref{Sec_model}~E.}
\label{fig_7}
\vskip -0.5cm
\end{figure*}
% ==============================================================================

Figures~\ref{fig_6} and \ref{fig_7} show the effect of this constraint on   various projections of the 
allowed parameter space.  We note that the classical expression (\ref{eq_alpha}) can be solved analytically 
for small $\alpha$, giving $|K|\!\approx \!0.9\Gamma$
for  $\Gamma'\!=\!\Gamma/2$, which agrees closely with the $\alpha\!=\!25^\circ$ boundary for the 
$K\!-\!\Gamma$ plane in Fig.~\ref{fig_6}, obtained numerically.
One can see that the most important effect of the constraint on the tilt angle is the lower boundary on $|K|$.
This is  physically meaningful as the tilt can only occur due to the non-zero anisotropic terms, $K$, $\Gamma$, and
$\Gamma'$, see also Ref.~\cite{kim19}.

Our second criterion is the  upper limit on the  energy bandwidth of the  magnetic intensity observed by the
neutron and Raman scattering \cite{nagler16,Banerjee17,kaib19,Wulferding19},
which sets a logical upper bounds on the  model parameters.
In  zero field, the bulk of the magnetic spectral weight is found below $\alt\!8$~meV, with 
extrapolations of the upper limit of the measurable signal extending to at most 15--20~meV.  

We make two general assumptions. First, the width of the detectable spectrum intensity 
is exhausted by one- and two-magnon excitations, which is true for the well-ordered phases such as the 
zero-field zigzag state of $\alpha$-RuCl$_3$. Even in the case of the pure Kitaev model, the basis of spin-flips 
is still complete, thus suggesting that this measure should provide a reasonable estimate of the bandwidth 
of any type of excitations. 
This yields the spectrum intensity width as $2W_0$, where $W_0$ is the ``bare'' LSWT single-magnon bandwidth.
Second, as is discussed above and in Sec.~\ref{Sec_quantum}, 
there is a quantum renormalization factor for the spectrum that can be approximated 
within the RPA \cite{Tyablikov} by a reduced  ordered moment, $\Lambda\!=\!\langle S\rangle/S$, thus narrowing 
the effective extent of the one- and two-magnon spectrum to $\approx\!2\Lambda W_0$.

The experimental estimates of the reduced ordered moment vary around $\Lambda\!\approx\! 0.5$ \cite{Coldea15,Cao16}, 
with the our  LSWT calculations in Sec.~\ref{Sec_quantum} suggesting a factor of 0.44. 
Taken together, the ``bare'' LSWT one-magnon bandwidth $W_0$ itself is roughly equivalent to an effective extent
of the detectable magnetic intensity. This consideration, together with the experimental limits discussed above,
create the basis for our  criterion. In the spirit of keeping this criterion ``soft'',  we present several versions of the 
constraint for $W_0$: ``realistic'' $\alt\!15$~meV, ``generous''  $\alt\!20$~meV, and ``outrageous'' $\alt\!25$~meV
cutoff values.

We also combine the constraint on  $W_0$  with the verification that the zero-field  ground state is indeed 
a zigzag state  for all parameter choices and that no intermediate phases occur between $H\!=\!0$ and $H\!=\!H_c$. 
For the first, we use the Luttinger-Tisza (LT) approach \cite{LT}, and for the second we inspect possible spectrum 
instabilities within the LSWT. This combination of the $W_0$ and zigzag  criteria is essential as they are 
less stringent separately.
 
Figures~\ref{fig_6} and \ref{fig_7} show the results of these constraints on different projections of the 
allowed parameter space.  The constraints on $W_0$ and zigzag are found numerically from the LSWT and LT, 
but for the boundaries shown in Figs.~\ref{fig_6} and \ref{fig_7} we use a fit \cite{W0_fit} that closely approximates them. 
One can see that the most important effect of these constraints is the upper boundary on $|K|$ and 
a  tighter bound on $\Gamma$ and  $J_{03}$ for the ``realistic''  $W_0$ limit.
%Since $K$  is expected to be the leading parameter of the model (\ref{eq_Hij}), this result is   natural.

For the $K$--$\Gamma$--$J_{03}$ subspace exhibited in Figs.~\ref{fig_6} and \ref{fig_7}, 
the boundary of the zigzag with an incommensurate  phase occurs at \emph{both} smaller and larger $|K|$.
For the smaller $|K|$, it is superseded  by the lower bound on the tilt angle $\alpha\!>\!25^\circ$. 
%It is the larger $|K|$ boundary that is  dependent on the choice of $W_0$. 
While the bandwidth limit does constrain the value of $|K|$ from above by itself, the combined effect 
with the zigzag requirement is considerably stronger. Thus, the  boundary of each color-coded shape 
in Figs.~\ref{fig_6} and \ref{fig_7} for large values 
of $|K|$ is also a boundary to an incommensurate state. This  is in a broad agreement with 
Ref.~\cite{NaglerVojta18}, where larger $|K|/|J|$  led to 
phases different from the zigzag, see also Appendix~\ref{app_0}.

\begin{figure}[t]
\centering
\includegraphics[width=0.99\linewidth]{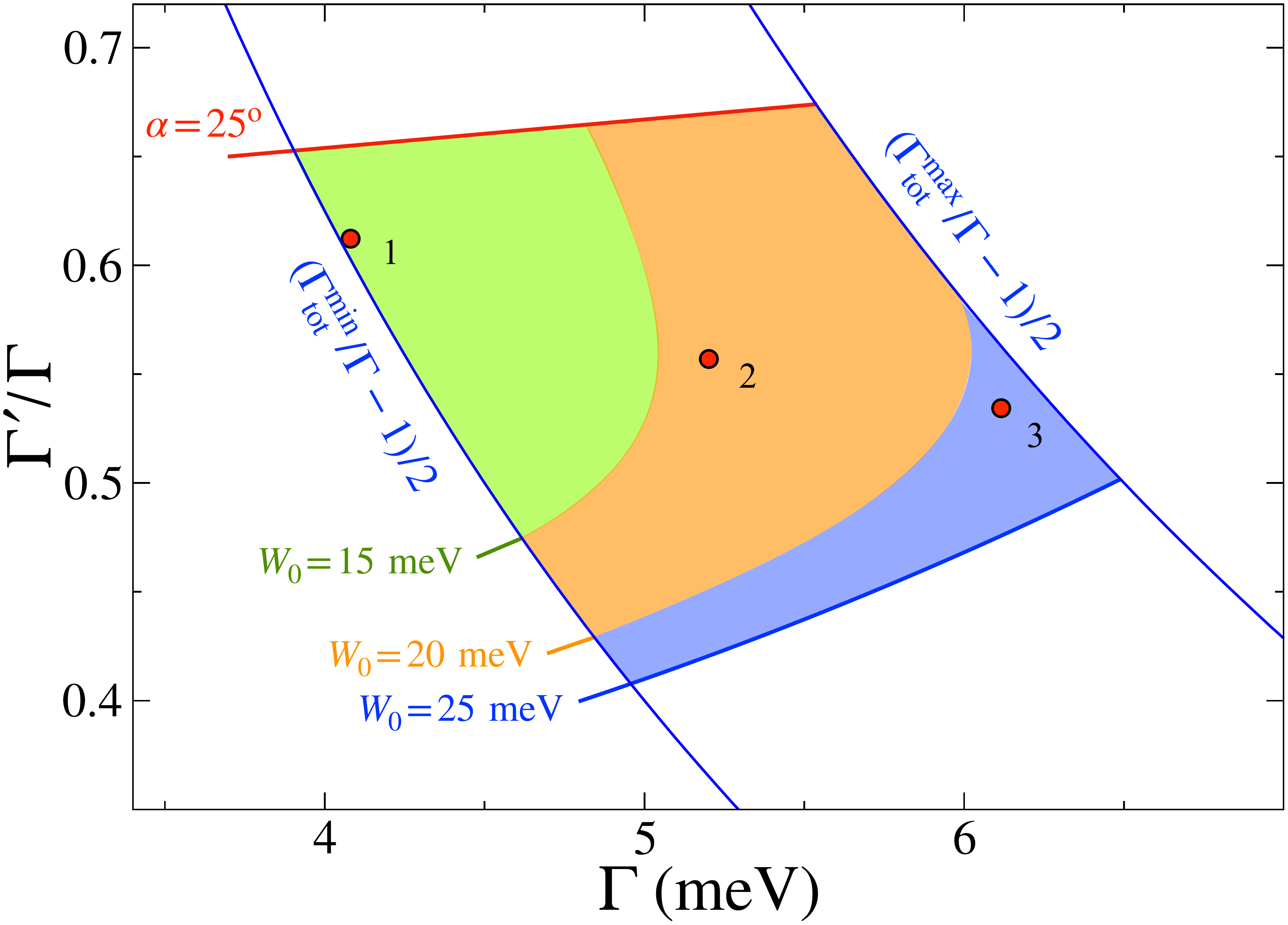}
\vskip -0.2cm
\caption{The $\Gamma$--$\Gamma'$ projection of the  parameter regions with all the constraints. 
Dots are  parameter sets from Sec.~\ref{Sec_model}~E.}
\label{fig_Gp}
\vskip -0.4cm
\end{figure}

In Figure~\ref{fig_Gp} we show a projection of the allowed parameter space onto 
the $\Gamma$--$\Gamma'$ plane, which quantitatively confirms our earlier assertion that $\Gamma'$ 
should be large and positive. The boundaries of the $\Gamma$--$\Gamma'$ region 
are set by the ESR/THz bounds on $\Gamma_{\rm tot}\!=\!\Gamma+2\Gamma'$, with 
$\Gamma^\text{min}_{\rm tot}\!=\!9$~meV and $\Gamma^\text{max}_{\rm tot}\!=\!13$~meV, as well
as by the lower boundary on the tilt angle $\alpha\!=\!25^\circ$ from above and on the bandwidth $W_0$
and zigzag from below. The latter are more stringent than the restrictions from $\Delta H_c$ (not shown), that were 
advocated  earlier. One can also observe that, overall, $\Gamma'$ is strongly tied to $\Gamma$, which, in turn,
is  $\approx\!\Gamma_{\rm tot}/2$.

% ==============================================================================
% ==============================================================================

\begin{figure}
\centering
\includegraphics[width=0.99\linewidth]{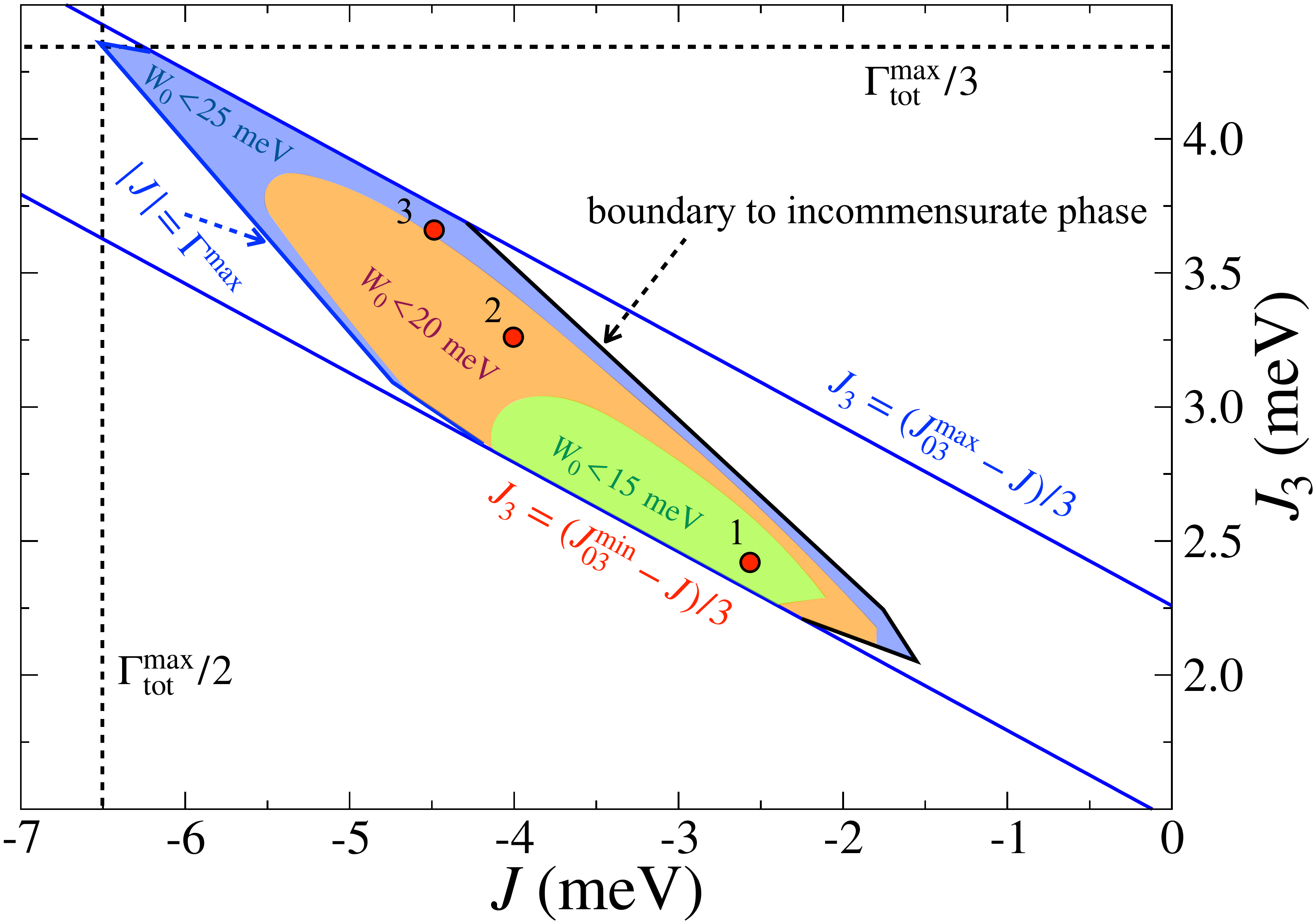}
\vskip -0.2cm
\caption{The $J$--$J_{3}$ projection of the parameter space, subject to all  constraints discussed in text. 
Boundaries to an incommensurate phase
are from the \textit{ab initio}-guided constraints $|J|\!<\!\Gamma^\text{max}\!=\!6.5$~meV, 
$|J|\!\alt\!\Gamma_{\rm tot}^\text{max}/2$,
and $J_3\!\alt\!\Gamma_{\rm tot}^\text{max}/3$, see text. 
Dots are  parameter sets from Sec.~\ref{Sec_model}~E.}
\label{fig_9}
\vskip -0.4cm
\end{figure}

Lastly, we present the $J$--$J_3$ projection in  Fig.~\ref{fig_9}.
According to our prior discussion, a combination of   $J$ and $J_3$,  termed
$J_{03}$ in (\ref{J03}), is  restricted and correlates narrowly with $\Gamma$, while, individually, 
these exchanges are unbounded.  In Figure~\ref{fig_9}, their allowed region is confined between 
parallel lines that are defined by the lower and  upper limits of  $J_{03}$  from Fig.~\ref{fig_7} and Eq.~(\ref{J03}).

The zigzag ground state is stabilized by the larger values of $|J|$ and $J_3$, in agreement 
with the previous work  that pointed out that trend  \cite{Kimchi11}. 
One can see that the constraints on the spectral width $W_0$ and on the zigzag state 
are responsible for the majority of the  $J$ and $J_3$ boundaries for both ``realistic'' 
and  ``generous'' $W_0$, while the parameter region for the ``outrageous'' (25~meV) choice of 
$W_0$  also encounters other boundaries.
 
These additional constraints are imposed by  the \textit{ab initio}  results.
As one can see in Table~\ref{table1}, the \textit{ab initio} methods set a rather strict hierarchy on 
the parameters of the model (\ref{eq_Hij}) of $\alpha$-RuCl$_3$:
$K$ and $\Gamma$ are dominant, while the rest of the terms are subleading. This produces an additional constraint,
$|J|,J_3\!<\!|K|,\Gamma$, that limits $|J|$ and $J_3$ from above. 
Since  $\Gamma$  is bounded by the ESR/THz constraints on $\Gamma_{\rm tot}$, this 
\textit{ab initio}-guided constraint together with the definition of $J_{03}$ in Eq.~(\ref{J03}) 
lead to close approximations for the upper limits on $|J|$ and $J_3$ in terms of 
$\Gamma^\text{max}_{\rm tot}$ shown in Fig.~\ref{fig_9}. In addition, one of the boundaries  
in Fig.~\ref{fig_9} is due to an explicit  constraint $|J|\!<\!\Gamma^\text{max}$.

% ==============================================================================
% ==============================================================================

\vspace{-0.3cm}

% ==============================================================================
\subsection{Summary}
% ==============================================================================
\vskip -0.2cm

Altogether, we have provided a series of empirical constraints  of varying rigidity on the parameters 
of the model (\ref{eq_Hij}) for $\alpha$-RuCl$_3$, with an overall qualitative, if  somewhat crude
take-home message. Most of the terms of the model are related to the same 
parameter, $\Gamma_{\rm tot}$, which is bounded by the ESR/THz constraints.
Specifically, $\Gamma \!\approx\! J_{03}\! \approx\! \Gamma_{\rm tot}/2$ and 
$\Gamma'\!\approx\!\Gamma/2$. The leading parameter is less constrained, with the  
ratio $|K|/\Gamma\!\approx\! 1.0-3.0$, and $J$ and $J_3$ require finer adjustments  to the experimental 
value of $H_c^{(a)}$. Empirically,  $|J|/\Gamma\! \approx\! J_3/\Gamma\! \approx\!0.5-0.75$.

These constraints and limits are represented in our Figs.~\ref{fig_6}, \ref{fig_7}, \ref{fig_Gp}, and \ref{fig_9}. 
In these Figures, we also show the representative points from each of the three  
regions bounded by the different constraint on the bandwidth $W_0$, with the following 
$(K,\Gamma,\Gamma',J,J_3)$ coordinates
and the corresponding values of $\{\Gamma_{\rm tot}, J_{03}\}$, all in meV:
\begin{center}
\begin{tabular}{rrrrrr | rl}
     & $(K,$ \ & $\Gamma,$   & $\Gamma',$  &  $J,$  & $J_3)\ $   &  $\{\Gamma_{\rm tot},$&$J_{03}\}$\\ 
{Point 1:} & (-4.8, & 4.08, & 2.5, & -2.56, & 2.42) \ & \    $\{9.08,$&$4.70\}$ \\ 
{Point 2:} & (-10.8, & 5.2, & 2.9, & -4.0, & 3.26) \ & \   $\{11.0,$&$5.78\}$\\ 
{Point 3:} & (-14.8, & 6.12, & 3.28, & -4.48, & 3.66) \ &  \  $\{12.7,$&$6.50\}$
\end{tabular}
\end{center}
Some  properties of these parameter sets, such as linear spin-wave spectrum, magnetization, N\'{e}el 
temperature, and critical fields, are presented in Sec.~\ref{Sec_quantum}.

\vspace{-0.3cm}
% ==============================================================================
\subsection{Compilation of parameters}
% ==============================================================================
\vskip -0.2cm

% ==============================================================================
\begin{table*}[t]
  \begin{tabular}{ l | c || c | c | c | c | c | c | c }
    \hline
\text{Reference} 
& \text{Method} 
& $K$                 & $\Gamma$ & $\Gamma'$ & $J$ & $J_3$ & $\Gamma\!+\!2\Gamma'$ & $J\!+\!3J_{3}$ \\ \hline\hline
Banerjee et al. \cite{nagler16} 
& LSWT, INS fit 
& +7.0                  &                  &                  & { -4.6}             &                  &               &     -4.6      \\ \hline 
\multirow{4}{2.5cm}{Kim et al. \cite{kee16}} 
& DFT+$t/U$, $P3$  
& \textbf{-6.55} & {\bf 5.25} & -0.95        & { -1.53}  &                  & 3.35       & -1.53       \\ \cline{2-9}
& DFT+SOC+$t/U$ 
& \textbf{-8.21} & {\bf 4.16} & -0.93        & -0.97           &                  & 2.3         & -0.97        \\ \cline{2-9}
& same+fixed lattice 
& -3.55                &  7.08         & -0.54         & {\bf -2.76} &                 & 6.01        & -2.76         \\ \cline{2-9}
& same+$U$+zigzag 
& +4.6                 & {  6.42} & -0.04         & {\bf -3.5}   &                  & 6.34        & -3.5            \\ \hline   
\multirow{2}{2.5cm}{Winter et al. \cite{winter16}} 
& DFT+ED, $C2$   
& \textbf{-6.67} & { 6.6}   & -0.87         & { -1.67} & {\bf 2.8}  & 4.87        & { 6.73}            \\ \cline{2-9}
& same, $P3$   
& +7.6                  & 8.4            &          +0.2 &            -5.5 & {\bf 2.3}   & {\bf 8.8} & +1.4              \\ \hline
Yadav et al. \cite{hozoi16} 
& Quantum chemistry 
& \textbf{-5.6}    &        -0.87 &                   &           +1.2 &                  &        -0.87 & +1.2              \\ \hline
Ran et al. \cite{wen17} & LSWT, INS fit 
& \textbf{-6.8}    &            9.5 &                  &                     &                  &  {\bf 9.5} &                    \\ \hline
&  DFT+$t/U$, $U\!=\!2.5$eV  
& { -14.43}     & { 6.43} &                  & {\bf -2.23}  & {\bf 2.07} & 6.43        &  +3.97 \\ \cline{2-9}
Hou et al. \cite{gong17} 
& same, $U\!=\!3.0$eV 
& { -12.23}     & {\bf 4.83} &                  & {\bf -1.93} &             1.6 &          4.83 &    +2.87        \\ \cline{2-9}
& same, $U\!=\!3.5$eV 
& {\bf -10.67}     & {\bf 3.8}   &                  & { -1.73} &           1.27 &            3.8 &   +2.07          \\ \hline
\multirow{2}{2.5cm}{Wang et al. \cite{li17}} 
& DFT+$t/U$, $P3$  
& {\bf -10.9}       & { 6.1}   &                  & -0.3              &           0.03 &            6.1 & -0.21            \\ \cline{2-9}
& same, $C2$  
& \textbf{-5.5}   & {7.6}          &                  & +0.1             &  0.1             &         7.6   & +0.4                 \\ \hline
Winter et al. \cite{winter17} 
& \textit{Ab initio}+INS fit 
& \textbf{-5.0}   &            2.5  &                  & -0.5              &              0.5 &            2.5 & +1.0                  \\ \hline
Suzuki et al. \cite{suga18} 
& ED, $C_p$ fit 
& -24.41              & {\bf 5.25} &         -0.95 & { -1.53}  &                    &           3.35 & -1.53               \\ \hline
Cookmeyer\! et\! al. \cite{moore18} 
& thermal Hall fit 
& \textbf{-5.0}   &            2.5 &                    &             -0.5 &           0.11 &             2.5  &     -0.16           \\ \hline
 Wu et al. \cite{orenstein18} 
& LSWT, THz fit 
& -2.8                  &            2.4 &                    &           -0.35 &           0.34 &              2.4 & +0.67                  \\ \hline
\multirow{2}{2.5cm}{Ozel et al. \cite{gedik19}} 
& same, $K\!>\!0$ 
& +1.15               & 2.92          & +1.27         & -0.95           &                    & 5.45           & -0.95                  \\ \cline{2-9}
& same, $K\!<\!0$ 
& -3.5                  & 2.35          &                    & +0.46          &                    & 2.35           & +0.46                    \\ \hline
Eichstaedt\! et\! al. \cite{berlijn19} 
& DFT+Wannier+$t/U$ 
& {-14.3} & 9.8            & -2.23         & -1.4             &           0.97 & 5.33            & +1.5                       \\ \hline
Sahasrabudhe\! et\! al.\cite{kaib19} & ED, Raman fit & \textbf{-10.0} & 3.75 & \textbf{} & -0.75 & 0.75 & 3.75 & 1.5 \\ \hline
Sears et al. \cite{kim19} 
& Magnetization fit 
& \textbf{-10.0} & 10.6          & -0.9           & {\bf -2.7}   &                    & {\bf 8.8}      & -2.7                    \\ \hline
Laurell et al. \cite{okamoto19} 
& ED, $C_p$ fit 
& {-15.1} & 10.1          & -0.12        & -1.3              &          0.9    & {\bf 9.86}  & +1.4                       \\ \hline\hline
\multirow{4}{2.cm}{\bf This work}  
%&  {\bf range}  \ \ 
%&\ {\bf  [-19,-3.8]}\ &\ {\bf [3.9,6.5]}\  &\ {\bf [2.0,3.6]}\ &\ {\bf [-6.5,-1.6]}\ &\ {\bf [2.1,4.3]}\ &\ {\bf [9.0,13.0]}\ &\ {\bf [4.4,6.8]} \\ 
&  {\bf ``realistic'' range}  \ \ 
&\ {\bf  [-11,-3.8]}\ &\ {\bf [3.9,5.0]}\  &\ {\bf [2.2,3.1]}\ &\ {\bf [-4.1,-2.1]}\ &\ {\bf [2.3,3.1]}\ &\ {\bf [9.0,11.4]}\ &\ {\bf [4.4,5.7]} \\
\cline{2-9}
&  {\bf point 1} 
& {\bf -4.8}          & {\bf 4.08} & {\bf 2.5}   &  {\bf -2.56} & {\bf 2.42}  & {\bf 9.08}  & {\bf 4.7}                \\ \cline{2-9}
& {\bf point 2}  
& {\bf -10.8}        & {\bf 5.2}   & {\bf 2.9}   & {\bf -4.0}    & {\bf 3.26}  & {\bf 11.0}   & {\bf 5.78}               \\ \cline{2-9}
&   {\bf point 3}
& { -14.8}        & { 6.12} & {\bf 3.28} & {  -4.48} & {  3.66}  & {  12.7}   &   {  6.5}                  \\ \hline
    \hline
\end{tabular}
\caption{The representative sets of parameters of the generalized KH model (\ref{eq_Hij}) for $\alpha$-RuCl$_3$ 
(in meV). The values that come close to the ranges proposed in this work are highlighted in bold.
The common acronyms include linear spin-wave theory (LSWT), density-functional theory (DFT),
spin-orbit coupling (SOC), inelastic neutron scattering 
(INS), exact diagonalization (ED), and terahertz spectroscopy (THz);
structures of $P3$ and $C2$ symmetry are referred to as ``$P3$'' and ``$C2$'' for brevity.}
\label{table1}
\vskip -0.2cm
\end{table*}
% ==============================================================================

Our Table \ref{table1} provides a representative compilation of the parameters of the model (\ref{eq_Hij})
that were proposed to describe $\alpha$-RuCl$_3$  
using the first-principles methods  \cite{winter16,kee16,gong17,li17,berlijn19,hozoi16} 
and phenomenological analyses 
\cite{wen17,winter17,suga18,moore18,orenstein18,gedik19,kim19,okamoto19,kaib19,nagler16}, with the first column 
providing the reference and the second abbreviating the details of the used approach. 
In cases when the proposed model description did not retain the $C_3$ symmetry of the ideal lattice structure,  
we used the bond-averaged values of the exchange parameters.

In Table \ref{table1}, we also present our results for the ranges of individual 
parameters for the ``realistic'' cutoff on $W_0$ and for the representative Point 1,2,3 sets described above. 
We note again,  that the parameters are correlated, with representative parameter sets illustrating these correlations. 
For instance, larger $\Gamma$ requires larger values of $\Gamma'$, $|J|$, etc. In that sense, 
the parameter ranges do not do full justice to the constraints that we advocate, as the actual 5-dimensional
constrained region is much narrower.

The listed values for each parameter are highlighted in bold in case they fall within or 
come close to our advocated ``realistic'' parameter range.
Although one can see quite a few ``hits'' in case of $K$, these may be mostly attributed to an extensive 
random shooting.  

There are two particularly notable differences  of our results from the prior studies.
First, is a significant and positive $\Gamma'$, which is either completely absent in the previous considerations
or is small and negative. In Section~\ref{Sec_model}, we have discussed extensively and made our case 
for the necessity of a significant $\Gamma'\!>\!0$ in the effective model description (\ref{eq_Hij}) of $\alpha$-RuCl$_3$.  

Second, are the ``cumulative'' parameters $\Gamma_{\rm tot}\!=\!\Gamma\!+\!2\Gamma'$ 
and  $J_{03}\!=\!J\!+\!3J_{3}$ in the last two columns of Table \ref{table1}. For the case of $\Gamma_{\rm tot}$,
there are a few studies providing comparable values, in which there is an attempt to describe phenomenology 
that is similar to ours, but without positive $\Gamma'$. These attempts 
can be seen as  trying to compensate for the lack of $\Gamma'$ by cranking up $\Gamma$ \cite{kim19,okamoto19}.
For $J_{03}$, it appears that previous works have, generally, underappreciated the importance of the mutual 
correlation of $J$ and $J_3$, leading to a nearly random distribution of their values.
As is discussed above, this work underscores the phenomenological 
constraints on  both $\Gamma_{\rm tot}$ and $J_{03}$ and the associated strong mutual bounds on 
$\Gamma$, $\Gamma'$, and $J_{03}$, see Figs.~\ref{fig_7} and \ref{fig_Gp}.

Lastly, as is emphasized in Sec.~\ref{Sec_intro} and Sec.~\ref{Sec_model},
the results of our work may differ from the prior analyses in Table~\ref{table1} because we consider phenomenology 
of an effective model as opposed to the first-principles methods, 
and we also extract bare parameters that are typically larger than the ones 
reduced by  quantum fluctuations.

\vspace{-0.3cm}
% ==============================================================================
\section{Quantum effects}
\label{Sec_quantum}
% ==============================================================================
\vskip -0.2cm

In this section,  we present the RPA results for the spectrum renormalization in the 
paramagnetic phase and demonstrate their close agreement with the ESR and THz data. 
As is shown above, the $\alpha$-RuCl$_3$ model has 
strong anisotropic-exchange interactions. They should inevitably lead to significant nonlinear quantum effects in the 
magnon spectra due to  substantial  three-particle interactions. Below, we calculate the damping of 
magnons due to associated decays and consider their effect in the spectrum  and  the dynamical structure factor.

\vspace{-0.3cm}

% ==============================================================================
\subsection{Quantum fluctuations}
\label{Sec_fluct}
% ==============================================================================
\vskip -0.2cm

% ==============================================================================
\begin{figure}
\centering
\includegraphics[width=0.99\linewidth]{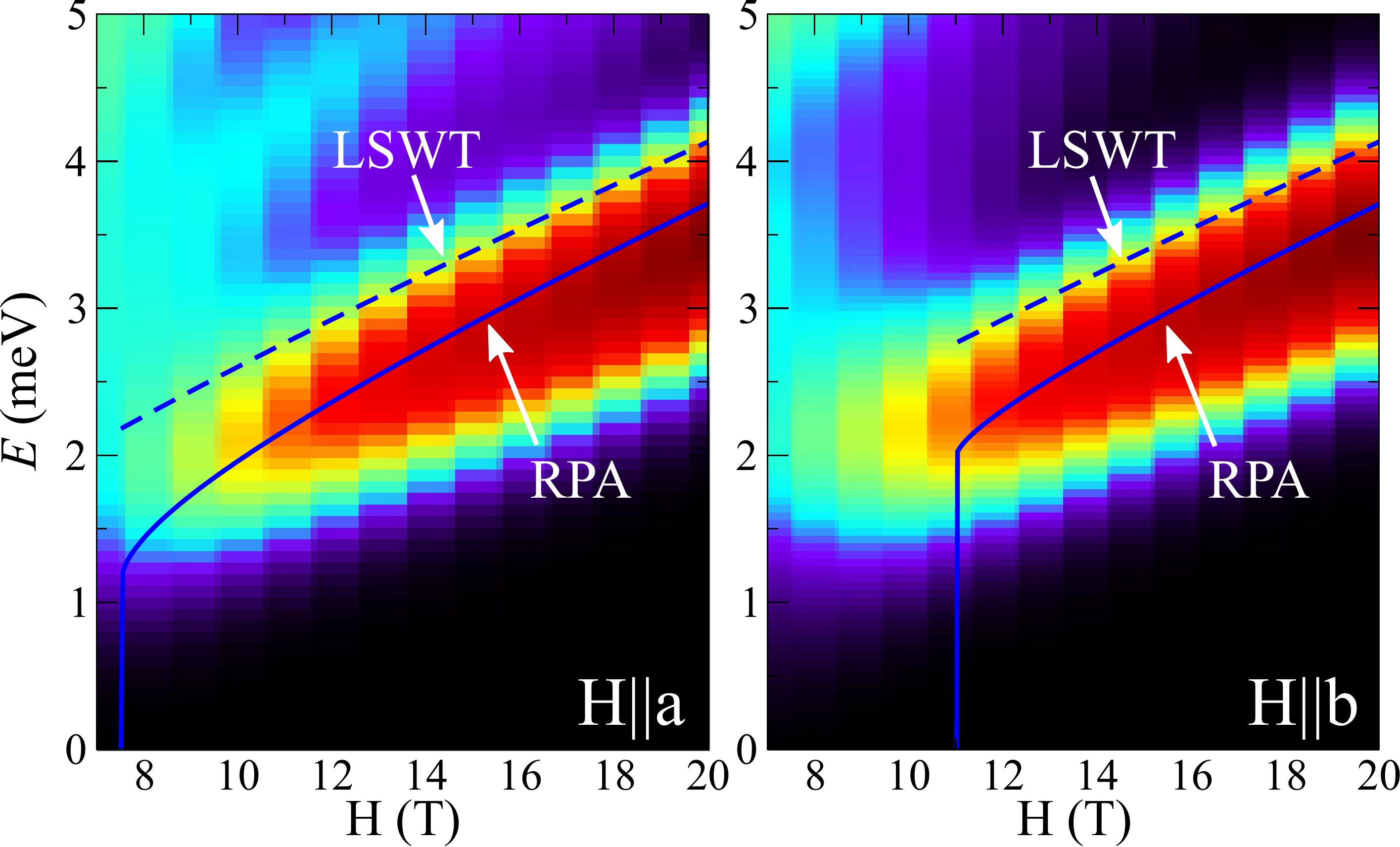}
\vskip -0.2cm
\caption{The ED results for the magnon spectrum at $\mathbf{q}\!=\!0$  as a function of magnetic field for 
two principal directions (intensity plot), reproduced from Ref.~\cite{Winter18}. Results  for the single-magnon 
branch by LSWT, Eq.~(\ref{Ek0}), (dashed line) and by the RPA (solid line) 
for the parameter set from Ref.~\cite{Winter18}.}
\label{fig_model2}
\vskip -0.2cm
\end{figure}
% ==============================================================================

The self-consistent RPA approach is based on the mean-field decoupling of the 
equations of motion for the spin Green's functions  \cite{Tyablikov,Plakida}. 
It provides an approximate, yet  effective, way of accounting for the downward spectrum renormalization
by quantum fluctuations. For $S\!=\!1/2$, the result is particularly simple \cite{Tyablikov,Plakida}
\begin{align}
\widetilde{\varepsilon}_\mathbf{k} = \Lambda \varepsilon_\mathbf{k}, \quad\Lambda=\langle S \rangle /S,
\label{eq_rpa}
\end{align}
where $\langle S \rangle$ is the average on-site magnetization reduced by quantum fluctuations, 
$\varepsilon_\mathbf{k}$ is the LSWT magnon energy, and $\widetilde{\varepsilon}_\mathbf{k}$ is the 
renormalized energy.

One can justify this approximation using  unbiased numerical methods. 
In Figure~\ref{fig_model2}, we provide a comparison of the RPA results with the 
ED calculations for the magnon energy spectrum at ${\bf q}\!=\!0$ in the paramagnetic phase 
vs field for two  field orientations. The ED data are from 
Ref.~\cite{Winter18} for the parameter set $K\!=\!-5$~meV, $\Gamma\!=\!2.5$~meV, 
$\Gamma'\!=\!0$,  and $J_3\!=\!-J\!=\!0.5$~meV, which has also been used in Refs.~\cite{winter17,moore18}
for different phenomenologies of  $\alpha$-RuCl$_3$.
One can see that RPA provides a significant improvement over 
the LSWT results and yields  a good agreement with the numerics.  

% ==============================================================================
\begin{figure}
\centering
\includegraphics[width=0.99\linewidth]{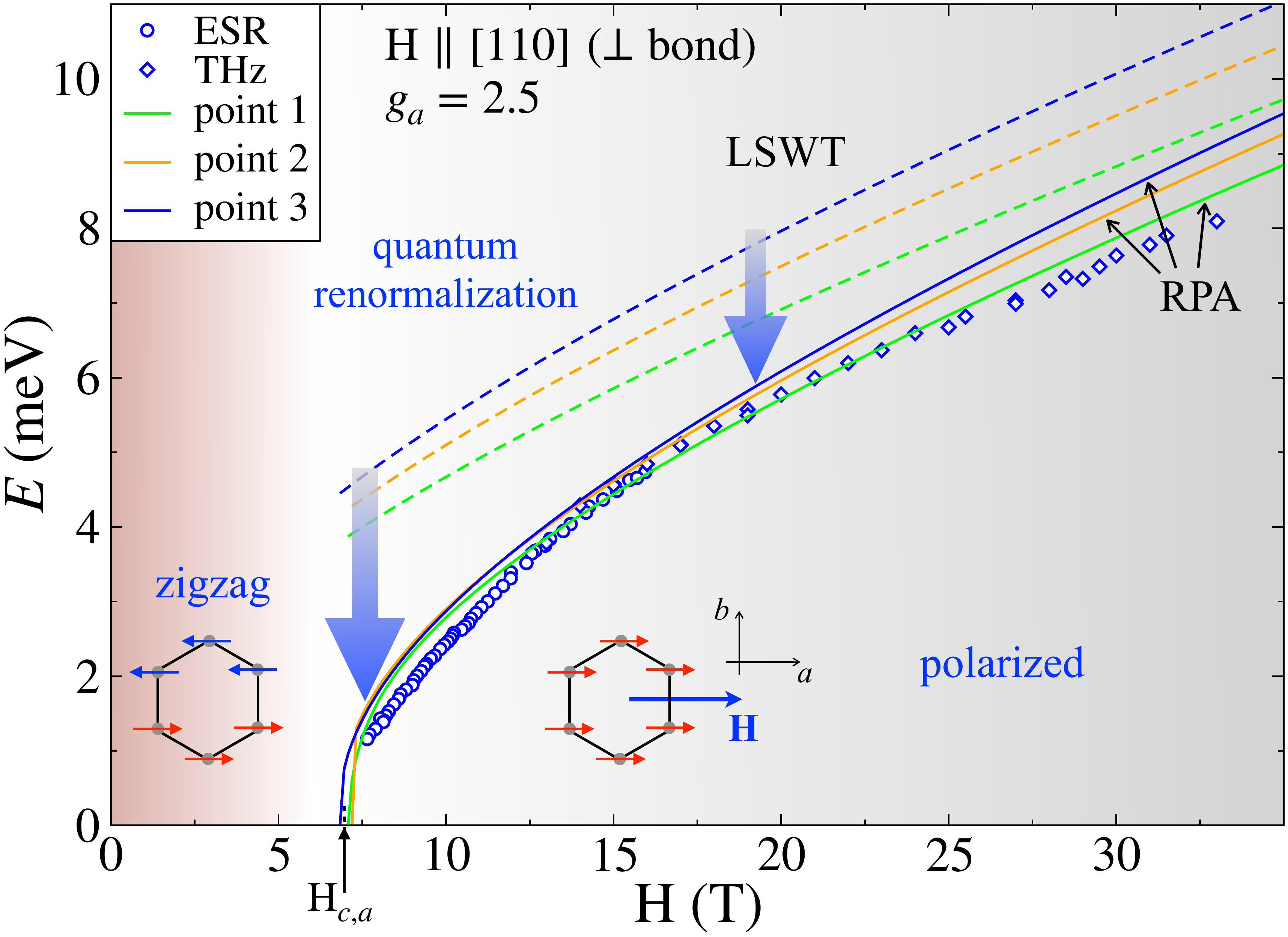}
\vskip -0.2cm
\caption{ESR  and THz   data for the single-magnon energy gap at ${\bf q}\!=\!0$ 
vs field from Fig.~\ref{fig_ESRfit} together with the LSWT (dashed lines) and RPA  (solid lines) results  
for the  representative points advocated in Sec.~\ref{Sec_model}~E and Table~\ref{table1}.}
\label{fig_3rpa}
\vskip -0.4cm
\end{figure}
% ==============================================================================

With this justification,  we now demonstrate the effect of %the downward renormalization by 
quantum fluctuations on the LSWT results for the single-magnon energy gap at ${\bf q}\!=\!0$  that was 
anticipated in Fig.~\ref{fig_ESRfit}. The results in Figure~\ref{fig_3rpa} are shown 
for the three  representative parameter sets, referred to as Point 1, 2, and 3 
in Sec.~\ref{Sec_model}~E and Table~\ref{table1}.
For all three sets, the mean-field RPA  already yields a close quantitative description of the ESR/THz data,
with the  Point 1 set, which belongs to the ``realistic'' region of the advocated parameter space,
giving the best fit of the three.

The variation of the slope of $\varepsilon_0(H)$ near the critical point has been attributed  
to changes of an effective $g$-factor and novel excitations \cite{Zvyagin17}. 
In Ref.~\cite{kaib19}, this effect has been ascribed to stronger repulsion from the two-magnon continuum 
in this field regime, which was also supported by the ED calculations. 
Here we corroborate the latter interpretation using the RPA method, which also shows a significant change of the slope 
due to enhanced quantum effects near the transition field. In our case,
in addition to the field dependence of the LSWT results from Eq.~(\ref{Ek0}),
an extra curvature of $\varepsilon_0(H)$ is due to the field dependence of  the 
ordered moment $\langle S \rangle$.

% ==============================================================================
\begin{table}[b]
  \begin{tabular}{ l | c | c | c | c | c }
  \hline
     & \ $\langle S \rangle$ \ & \ $T_{\rm MF}$, K \ & \ $T_N$, K \ &\  $H^{(a)[(b)]}_c$, T \ &\ \ $\alpha$,$^{\degree}$ \ \\ \hline
  Point 1\  &\  0.219 \ & 16.0 & 12.3 & 7.14 [7.88] & \ 28.2\ \\ \hline
  Point 2 \ &\ 0.220 \ & 25.4 & 16.2 & 7.20 [7.87] & \ 37.2\ \\ \hline
  Point 3 \ &\ 0.225 \ & 31.1 & 21.5 & 6.94 [7.79] & \ \ 39.4 \ \\ \hline
\end{tabular}
\caption{Zero-field magnetization, N\'{e}el temperature, critical fields, and tilt angle for representative 
 parameter sets.}
  \label{table_app}
\end{table}
% ==============================================================================

We   summarize some of the  zero-field properties of the model (\ref{eq_Hij})  
for the Point 1, 2, and 3 parameter sets
in Table~\ref{table_app}, where we present spin-wave results for the ordered moment, 
N\'{e}el temperature, critical fields from 
Eqs.~\eqref{Hca} and \eqref{Hcb}, and tilt angle from Eq.~(\ref{eq_alpha}) for all three sets.

Our LSWT calculations yield ordered moments that are indicative of strong fluctuations, 
$\langle S \rangle\!\approx\! 0.22$, the value that is in agreement with experimental 
estimates \cite{Coldea15,Cao16} and is also similar to the results for 
the Heisenberg antiferromagnet on the same lattice \cite{Weihong91}. 
While LSWT calculation of the ordered moment $\langle S \rangle$ is standard,  
the N\'{e}el temperature is calculated from a self-consistent condition 
on the ordered moment $\langle S \rangle\!\rightarrow\! 0$ at $T\!\rightarrow\!T_N$ 
within the spin Green's function formalism using RPA, 
see  Refs.~\cite{Zhu19,Plakida,Tyablikov} and Appendix~\ref{app_TN} for details.
One can see a significant lowering of the mean-field results for the ordering temperature,
with the latter obtained  from $T_{\rm MF}\!=\!-S(S+1)  \lambda_{\rm min} (\mathbf{Q})/3 k_B$,
where $\lambda_{\rm min} (\mathbf{Q})$ is the lowest eigenvalue of the Fourier transform of the 
exchange matrix in (\ref{eq_Hij})  at the ordering vector $\mathbf{Q}$ \cite{Gingras}. 
The experimental value of $T_N$ for $\alpha$-RuCl$_3$ is known to be sensitive to the stacking of the 
honeycomb planes \cite{Cao16} and can be lower than our values in Table~\ref{table_app}, which is likely  
related to the frustrating 3D interplane couplings \cite{Vojta3D}. 

As was discussed above, the critical fields are not changed by   quantum effects within the RPA,
as,  at the mean-field level, the effect of fluctuations  on the field-induced spin polarization cancels the same effect 
on the gap that is closing at the transition. While this is a mean-field argument, it points to suppressed 
quantum effects on the critical fields. In Table~\ref{table_app},  the critical fields are from 
Eqs.~\eqref{Hca} and \eqref{Hcb}, and for all three sets they are  close to the experimental values for 
$\alpha$-RuCl$_3$ \cite{NaglerVojta18}.
Altogether, the RPA approach provides a good overall description of several aspects of 
$\alpha$-RuCl$_3$ phenomenology.

The zero-field ordered moment $\langle S \rangle$  in Table~\ref{table_app} is  about the same
for all three representative parameter sets, the feature that can be attributed to a significant similarity of the 
spin-wave spectra in all three cases shown in Fig.~\ref{fig_lswt3} for a ${\bf q}$-contour through the Brillouin zone,
see Fig.~\ref{fig_strfac} below.
The spectra consists of four branches due to the four-sublattice structure of the zigzag state.  
Since  the Point 1, 2 and 3  sets belong to the parameter regions with  different bandwidth limits, this provides  
the main difference between otherwise  similar plots. 

We also note the pseudo-Goldstone mode at the M 
point in all three plots that occurs due to an accidental degeneracy. 
The nature of this degeneracy will be discussed in Sec.~\ref{Sec_duality}. 
The experimental value of the gap at the M point  is larger 
than in our  Fig.~\ref{fig_lswt3}, which is related to the $C_3$ symmetry breaking in $\alpha$-RuCl$_3$
\cite{kaib19,winter16}. While strongly affecting the gap at the accidental degeneracy points, this
lower symmetry is not expected to significantly change other results discussed in this work \cite{footnote_C3}.

% ==============================================================================
\begin{figure}
\centering
\includegraphics[width=0.99\linewidth]{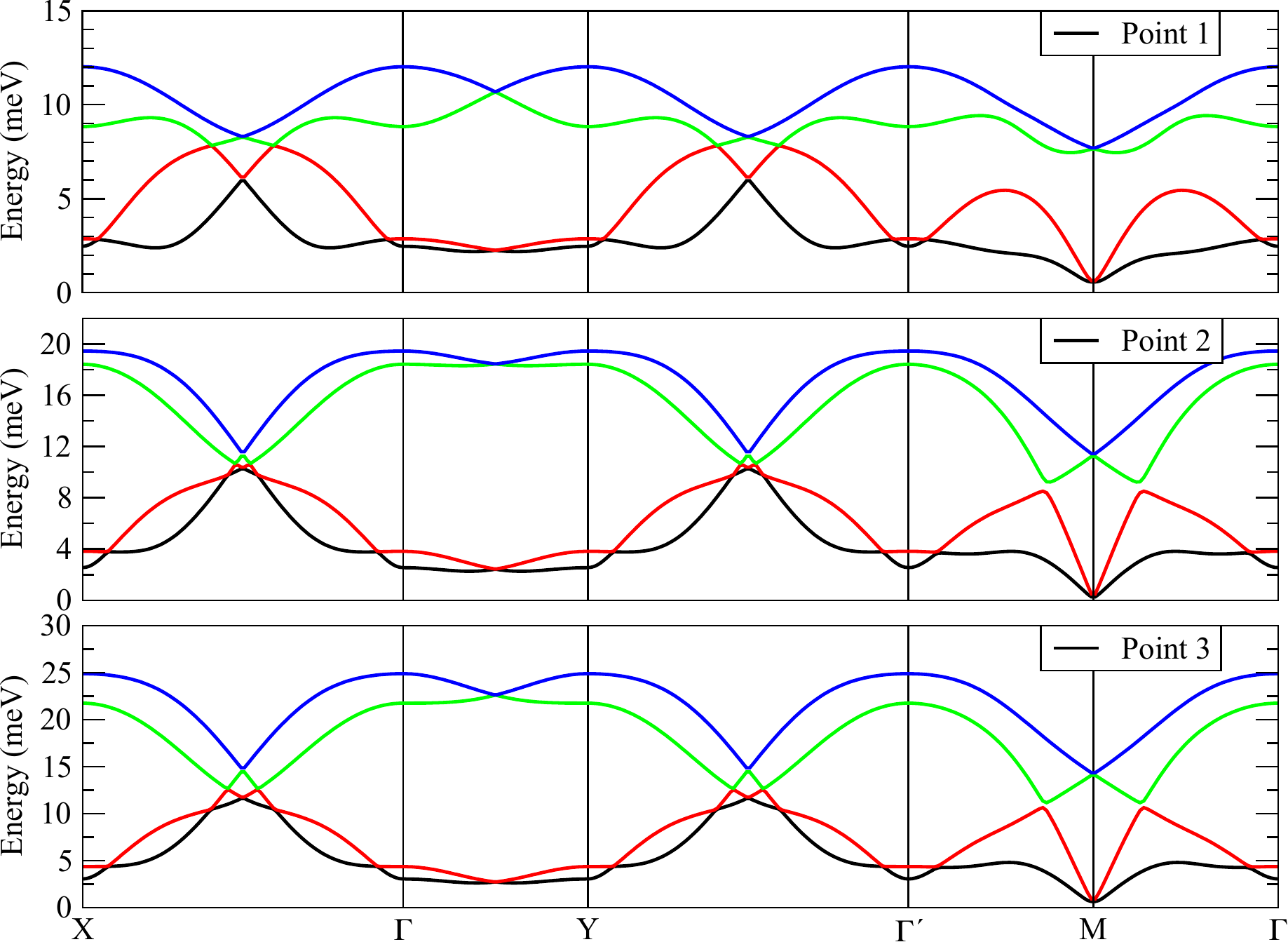}
\vskip -0.2cm
\caption{The LSWT magnon spectra for three representative parameter sets, Point 1, 2, and 3 from
Sec.~\ref{Sec_model}~E and Table~\ref{table1}.}
\label{fig_lswt3}
\vskip -0.4cm
\end{figure}
% ==============================================================================

\vspace{-0.3cm}

% ==============================================================================
\subsection{Magnon decays}
\label{Sec_decays}
% ==============================================================================
\vskip -0.2cm

It was argued in Refs.~\cite{winter17, kopietz20} that strong  quantum effects 
are expected to be generally present in the spectrum of any anisotropic-exchange magnet, 
except in some narrow regions of its phase diagram where the off-diagonal exchange terms 
are suppressed. The significant off-diagonal terms \emph{necessarily} produce strong anharmonic couplings of magnons 
for any form of the underlying magnetic order. Such couplings, in turn, inevitably lead to a nearly complete 
wipeout of the  higher-energy magnons  due to large decay  rates into the lower-energy magnon continua \cite{RMP13}.  
As a result, much of the magnon spectrum observed by the inelastic neutron scattering is expected to be 
comprised of the broad features combined with some well-defined low-energy magnon modes. 

Since the preceding consideration unequivocally points to large off-diagonal $\Gamma$ and $\Gamma'$ terms 
in the model (\ref{eq_Hij}) of $\alpha$-RuCl$_3$, there is no question in our mind that the scenario 
advocated in  Refs.~\cite{winter17, kopietz20} is applicable here. 
Below, we first repeat the general arguments for the inevitability of strong magnon decays, and then 
apply the approximate analysis of  them to a representative set of the model parameters
and demonstrate a coexistence  of the well-defined low-energy quasiparticles with the 
broadened excitation continua. We argue that these results are in agreement with the experimental features 
found in the spectrum of $\alpha$-RuCl$_3$ \cite{nagler16,wen17,Banerjee17,banerjee18}.
We underscore, once again, the importance of taking into account magnon decays
in interpreting broad features in the spectra of   all strongly-anisotropic magnets \cite{JeffYbTiO}. 

\vspace{-0.3cm}

% ==============================================================================
\subsubsection{General formalism}
% ==============================================================================
\vskip -0.2cm

Within the   spin-wave expansion, the  reference frame on each site 
is rotated to a local one  with the new $z$ axis pointing along the spin's quantization axis that is 
given by the classical energy minimization, $\mathbf{S}_i\! =\!\hat{\bf R}_i  \widetilde{\mathbf{S}}_i$.
Here, $\widetilde{\mathbf{S}}_i$ is the spin vector in the local reference frame at the site $i$ and $\hat{\bf R}_i$
is the rotation matrix for that site.
For the case of the model (\ref{eq_Hij}) of $\alpha$-RuCl$_3$, this would be a rotation from the cubic axes 
to the   axes of the zigzag order that are tilted out of the basal plane of the honeycomb lattice.

Thus, the spin Hamiltonian \eqref{eq_Hij}  can be rewritten as 
\begin{align}
\hat{\cal H}=\sum_{\langle ij\rangle} \widetilde{\mathbf{S}}_i^{\rm T} 
\widetilde{\bm J}_{ij} \widetilde{\mathbf{S}}_j\, ,
\label{eq_Hij_loc}
\end{align}
where the ``rotated'' exchange matrix is
\begin{align}
\widetilde{\bm J}_{ij} =\hat{\bf R}_i^{\rm T}\hat{\bm J}_{ij}\hat{\bf R}_j
=\left(
\begin{array}{ccc}
\widetilde{J}_{ij}^{xx} &\widetilde{J}_{ij}^{xy} &\widetilde{J}_{ij}^{xz} \\
\widetilde{J}_{ij}^{yx} &\widetilde{J}_{ij}^{yy} &\widetilde{J}_{ij}^{yz} \\
\widetilde{J}_{ij}^{zx} &\widetilde{J}_{ij}^{zy} &\widetilde{J}_{ij}^{zz} 
\end{array} \right).
\label{eq_Jij_rot}
\end{align}
For the model (\ref{eq_Hij}) one can ignore the third-neighbor Heisenberg exchange terms because 
they do not contribute to decays for a collinear zigzag order.

The LSWT needs only   diagonal and $\widetilde{J}_{ij}^{xy}(\widetilde{J}_{ij}^{yx})$ terms of the matrix 
$\widetilde{\bm J}_{ij}$, while it is its off-diagonal parts that give rise to the three-magnon interaction
\begin{align}
\label{H3}
\mathcal{H}_{\rm od}=\sum_{\langle ij \rangle}\left( \widetilde{J}_{ij}^{xz}
\widetilde{S}^x_i \widetilde{S}^z_j +\widetilde{J}_{ij}^{yz} 
\widetilde{S}^y_i \widetilde{S}^z_j +\{i\leftrightarrow j\}\right).
\end{align}
The form of the ``original'' exchange matrix ${\bm J}_{ij}$ in (\ref{H_JKGGp}),
in which all terms of the generalized KH model are present, 
makes it clear that it is only some very restrictive choices of the  parameters and ordered states that
can render the resultant off-diagonal $\widetilde{J}_{ij}^{xz}$ and $\widetilde{J}_{ij}^{yz}$ terms negligible. 

For the  zigzag state within the model (\ref{H_JKGGp}), we obtain 
explicit expressions of $\widetilde{J}_{ij}^{xz}$ and $\widetilde{J}_{ij}^{yz}$  for each non-equivalent bond
 in terms of $K$, $\Gamma$,  $\Gamma'$, 
and polar and azimuthal angles  of the zigzag axes relative to the cubic ones.
They are listed in Appendix~\ref{app_B}. As was discussed in Ref.~\cite{winter17}, 
the off-diagonal couplings in Eq.~(\ref{H3}) are non-zero except for the 
case $\Gamma\!=\!\Gamma'\!=\!0$, which also makes spins orient along one of the cubic
axes. Given that in case of $\alpha$-RuCl$_3$, 
$K$ and $\Gamma$ are, in fact, the leading terms of the generalized KH model, it is natural  
that the off-diagonal couplings $\widetilde{J}_{ij}^{xz}$ and $\widetilde{J}_{ij}^{yz}$ are very significant. Thus, it is 
imperative to consider their effect in the spin-wave excitations. 

The Holstein-Primakoff bosonization of Eq.~\eqref{H3} yields the three-particle interaction 
\vskip -0.05cm
\noindent
\begin{align}
\label{H3HP}
\mathcal{H}_3= \sum_{\langle ij \rangle} \widetilde{V}_{ij} 
\left( a^\dagger_i a^\dagger_j a^{\phantom \dagger}_j+\text{H.c}+\{i \rightarrow j\}\right),
\end{align}
\vskip -0.1cm
\noindent
where $\widetilde{V}^{}_{ij}\!=\! -\sqrt{S/2} \left( \widetilde{J}_{ij}^{xz}+i\widetilde{J}_{ij}^{yz}\right)$.

While the technical procedure of obtaining  fully symmetrized three-magnon interactions  
from the Holstein-Primakoff form of Eq.~(\ref{H3HP}) typically requires numerical 
diagonalization of the quadratic LSWT Hamiltonian and is also quite involved otherwise, see Ref.~\cite{kopietz20},
the resultant form of the decay part of it is general,
\begin{eqnarray}
\hat{\cal H}_3\!=\!\frac{1}{2\sqrt{N}}\!\sum_{{\bf k}+{\bf q}=-{\bf p}}\sum_{\eta\nu\mu} \left(
\widetilde{V}^{\eta\nu\mu}_{{\bf q}{\bf k};{\bf p}} 
d^{\dagger}_{\eta{\bf q}} d^{\dagger}_{\nu{\bf k}} d^{\phantom{\dag}}_{\mu-{\bf p}}+{\rm H.c.}\right)\!, \ \ \ \ 
\label{Hdecay}
\end{eqnarray}
where $d^{\dagger} (d)$ are magnon operators, indices $\eta$, $\nu$, and $\mu$ numerate magnon 
branches, and $\widetilde{V}^{\eta\nu\mu}_{{\bf q}{\bf k};{\bf p}}$ is the vertex.
With this interaction (\ref{Hdecay}),  standard diagrammatic rules allow for a systematic calculation
of the quantum corrections to the magnon spectra in the form of  self-energies $\Sigma^\mu (\mathbf{k},\omega)$. 

\vspace{-0.3cm}

% ==============================================================================
\subsubsection{Approximations}
% ==============================================================================
\vskip -0.2cm

The standard approach, justified within the $1/S$ expansion, is to consider a one-loop correction to the
spectrum due to three-magnon interaction (\ref{Hdecay}). Since the most drastic qualitative effect of decays
is the finite lifetime, one can ignore the real part of the self-energy of the branch $\mu$ 
and calculate it in the on-shell approximation, 
$\Sigma^\mu (\mathbf{k},\omega)\!\approx\!-i\Gamma^{\mu}_{{\bf k}}$, 
where the decay rate of the mode $\mu$ is 
\begin{eqnarray}
\Gamma^{\mu}_{{\bf k}} =
\frac{\pi}{2N} \sum_{{\bf q},\eta\nu} 
\big| \widetilde{V}^{\eta\nu\mu}_{{\bf q},{\bf k-q};{\bf k}}\big|^2 
\delta(\varepsilon_{\mu{\bf k}}-\varepsilon_{\eta{\bf q}}-\varepsilon_{\nu{\bf k-q}}).
\label{eq_gammak}
\end{eqnarray}
Below we will also capitalize on the apparent success of the RPA approach to account for the renormalization
of the real part of magnon energies in a simplified fashion.

Generally, the use of Eq.~(\ref{eq_gammak}) together with the derivation of the vertex in Eq.~(\ref{Hdecay}) 
requires numerical diagonalization of the LSWT Hamiltonian and matrix transformations with 
potentially prohibitive  computational costs. Instead, we use the 
``constant matrix element'' approach that was proposed in Ref.~\cite{winter17} and was recently
validated for the generalized KH  model  in Ref.~\cite{kopietz20},
were it was found to provide a good quantitative  approximation. 
We briefly describe its  nature below.

The decay rate (\ref{eq_gammak}) can be related to a  simpler quantity, 
the on-shell two-magnon density of states (DoS),
 \begin{align}
D_\mathbf{k}(\varepsilon_{\mu\bf k})
=\frac{\pi}{N}\sum_{\mathbf{q},\nu\eta}\delta \left( \varepsilon_{\mu\mathbf{k}}-\varepsilon_{\nu\mathbf{q}}
-\varepsilon_{\eta\mathbf{k-q}}\right),
\label{eq_dos}
\end{align}
which quantifies the overlap of the single-magnon excitations  of the branch $\mu$
with the two-magnon continuum.

Since we have the full knowledge of the real-space three-magnon vertices in Eq.~(\ref{H3HP}), see Appendix~\ref{app_B},
we can introduce the overall strength of the coupling  
\begin{equation}
\widetilde{V}_{\rm eff}=\frac{1}{12} \sum_{i}\sum_{\langle ij \rangle_\gamma} \big|\widetilde{V}_{ij} \big|,
\label{eq:3_scale}
\end{equation}
where $i$  sums over four sublattices of the zigzag state  and $\gamma\!=\!\{{\rm X,Y,Z}\}$
is numerating the bonds. This definition is consistent with the ones used in Refs.~\cite{winter17,kopietz20}. 
Then, one can rewrite the three-magnon vertex in Eq.~(\ref{Hdecay}) as
\begin{eqnarray}
\widetilde{V}^{\eta\nu\mu}_{{\bf q}{\bf k};{\bf p}} =\widetilde{V}_{\rm eff}\, 
\widetilde{\Phi}^{\eta\nu\mu}_{{\bf q}{\bf k};{\bf p}},
\end{eqnarray}
where the  dimensionless vertices $\widetilde{\Phi}$ include all the necessary transformations and symmetrizations. 

Within the ``constant matrix element'' approximation, we substitute the dimensionless 
$\big| \widetilde{\Phi}^{\eta\nu\mu}_{{\bf q},{\bf k-q};{\bf k}}\big|^2$
in the decay rate (\ref{eq_gammak}) by a constant, thus eliminating the numerically costly and analytically 
cumbersome element of the calculation.
As a result, the decay rate (\ref{eq_gammak}) is simply proportional 
to the on-shell two-magnon DoS (\ref{eq_dos})
\begin{align}
\Gamma^\mu_\mathbf{k} \approx \frac{f}{2} \big|\widetilde{V}_{\rm eff}\big|^2 D(\varepsilon_{\mu\bf k}),
\label{eq_gamma_app}
\end{align}
with  an implied relation of the average
 $f\!=\!\langle\big| \widetilde{\Phi}^{\eta\nu\mu}_{{\bf q},{\bf k-q};{\bf k}}\big|^2\rangle$.
This approximation leads to a drastic simplification for the decay rate calculation, as one needs only 
magnon energies from the  harmonic theory and the average real-space three-magnon coupling strength 
from Eq.~(\ref{eq:3_scale}). 

One of the strong justifications of the constant matrix element approximation 
is the common origin of the Van Hove singularities in the decay rates and the two-magnon DoS. 
This approximation is also significantly improved by using the self-consistency
within the Dyson's equation, referred to as the iDE approach 
\cite{winter17,kopietz20,tri09,tri_H,FM_DM},  
\begin{align}
\Gamma^{\mu}_{{\bf k}} = \frac{f}{2} \big|\widetilde{V}_{\rm eff}\big|^2 D^{\mu}_{{\bf k}}
(\varepsilon_{\mu\mathbf{k}}+i\Gamma^{\mu}_{{\bf k}}),
\label{eq:cv_ide}
\end{align}
where the $\delta$-function with the complex argument is a shorthand for a Lorentzian.
Since within the iDE approach there is an effective averaging over various states, 
it provides further credence  to the  constant matrix element approximation. 

%----------------------------------------------------------------------------
\begin{figure}[t]
\includegraphics[width=.99\linewidth]{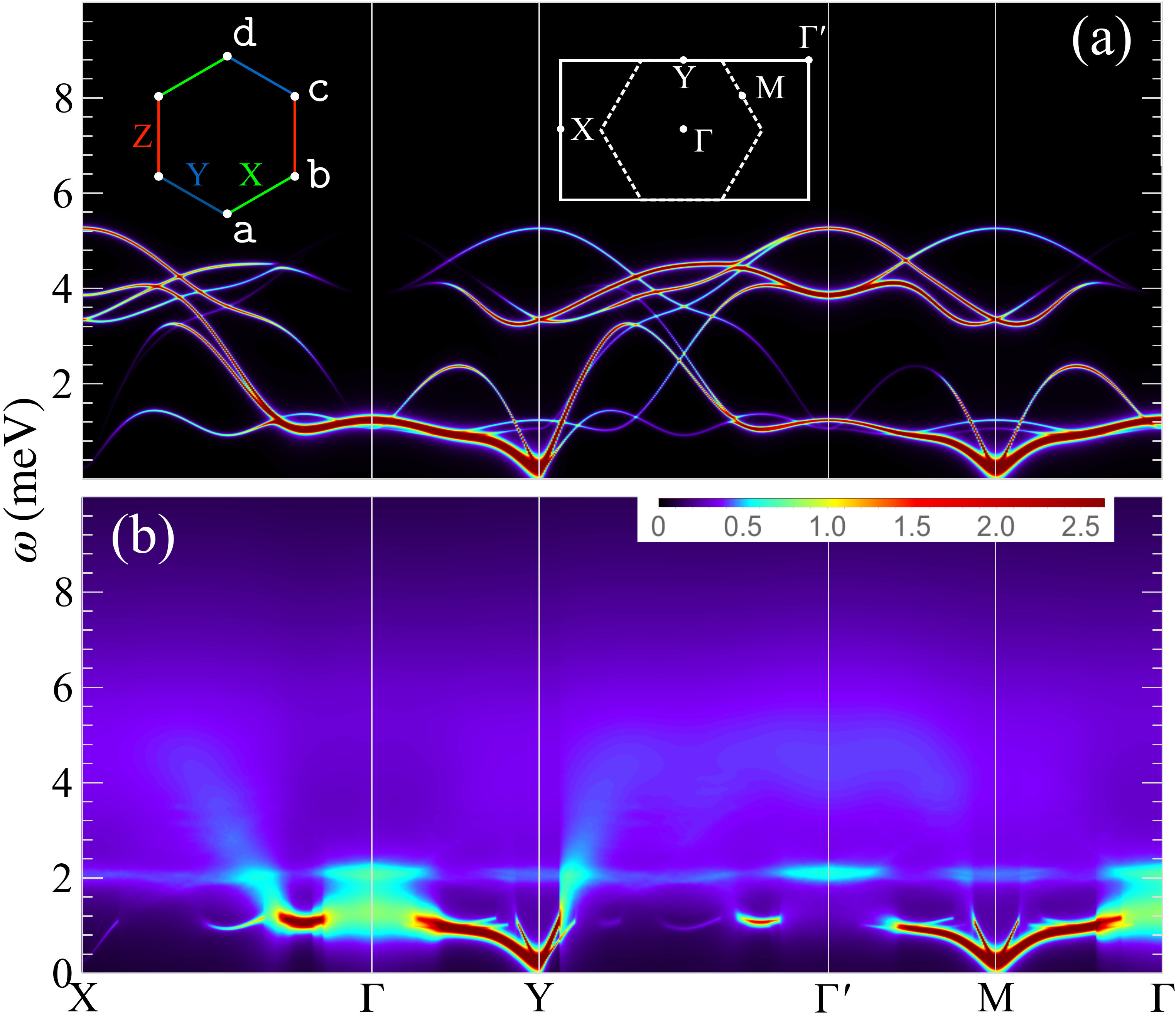} 
\vskip -0.2cm
\caption{The dynamical structure factor ${\cal S}({\bf q},\omega)$  
for the Point~1 parameter set: (a) LSWT results with RPA renormalization, (b) includes  
magnon lifetime effects and  longitudinal contribution, see text.}
\label{fig_strfac}
\vskip -0.4cm
\end{figure}
%----------------------------------------------------------------------------

\vspace{-0.3cm}

% ==============================================================================
\subsubsection{Results}
% ==============================================================================
\vskip -0.2cm

In Figure~\ref{fig_strfac}, we present the results of the calculation of the dynamical structure factor 
${\cal S}({\bf q},\omega)$ for the representative parameter set of Point 1 from 
Sec.~\ref{Sec_model}~E and Table~\ref{table1}.
Fig.~\ref{fig_strfac}(a) shows the LSWT results with all energies multiplied by the RPA renormalization factor,
$\langle S \rangle/S \approx 0.44$, argued for above, see also Table~\ref{table_app}. Since within the LSWT  
only transverse component of the structure factor contributes, this panel  
showcases  the mean-field renormalization effect of  quantum fluctuations on the bare 
spectrum of Fig.~\ref{fig_lswt3}. We  note that the results in Fig.~\ref{fig_strfac} are averaged over three
equivalent domains of the zigzag order, the intensity is cut below the highest maxima in order to emphasize 
details of the structure factor, and  artificial broadening is $\delta\!=\!0.05$~meV.
 
Fig.~\ref{fig_strfac}(b) is our main result. It shows a mix of the contributions of the transverse and longitudinal components 
of the structure factor, both taking into account magnon lifetime effects on top of the RPA renormalization. 
Compared to Fig.~\ref{fig_strfac}(a), the transverse component  includes the decay rates obtained 
within the iDE approach (\ref{eq:cv_ide}).
For the Point 1 parameter set, the calculated overall strength of the three-magnon coupling (\ref{eq:3_scale}) is
$V_\text{eff}\!=\!6.1$~meV, which is about half of the bare magnon bandwidth in Fig.~\ref{fig_lswt3} for the same set.
This value is in accord with the  expectations of the significant off-diagonal terms 
in the exchange matrix laid out above.
For the {\it ad hoc}  parameter $f$ in the calculations of the damping in Eq.~(\ref{eq:cv_ide}), the choice 
was made at $f\!=\!0.1$, which is similar to the ones used in the prior works, see Refs.~\cite{winter17,kopietz20},
where it was based on a comparison with the fully microscopic calculations for the 
honeycomb-lattice $XXZ$ model in a field \cite{hex_H} and  for the KH-$\Gamma$ 
model \cite{kopietz20}. 

As a result of the damping, the upper magnon modes are strongly washed out, with the lower   
branches damped in some regions of the Brillouin zone and stable in the others. In particular, regions 
near the $\Gamma$ point are broadened due to decays into the quasi-Goldstone modes,
in an accord with the results of Ref.~\cite{winter17} for a different set of parameters. Overall, 
the broadening is  stronger in the present case because the three-magnon coupling is  
larger. 

An important element to the structure factor in the strongly fluctuating system is the
longitudinal component that probes the %comes from the two-magnon contribution, thus probing the 
two-magnon continuum directly. To keep its description on a par with that of the constant matrix element 
approximation for the decays into such a continuum, we approximate the longitudinal structure factor 
as directly proportional to the two-magnon density of states with the help of another {\it ad hoc} constant parameter, 
bypassing the need for cumbersome manipulations with the magnon eigenvectors
\begin{align}
\label{Szz}
\mathcal{S}^{zz}(\mathbf{q},\omega)=\frac{\pi f_2}{N}\sum_{\mathbf{k},\nu\eta}\delta \left( \omega-
\widetilde{\varepsilon}_{\nu\mathbf{k}}
-\widetilde{\varepsilon}_{\eta\mathbf{q-k}}\right),
\end{align}
where $\widetilde{\varepsilon}_{\nu\mathbf{k}}\!=\!\varepsilon_{\nu\mathbf{k}}+
i\Gamma^\nu_\mathbf{k}$ is the RPA-adjusted magnon energy of the mode $\nu$ together with its damping.
Including broadening effects in the magnon lines in Eq.~(\ref{Szz}) adds another level of self-consistency to our calculation.
Here, an improvement over   Ref.~\cite{winter17}  is the use
of the momentum-dependent $\Gamma^\nu_{\mathbf{k} ({\bf q-k})}$ instead of the averaged ones,  resulting
in more pronounced Van Hove singularities in the two-magnon continuum.
The choice  $f_2\!=\!2/15$ is also similar to the prior work \cite{winter17}.

In Figure~\ref{fig_strfac}(b), the two-magnon longitudinal component provides a strong contribution to the signal at 
higher energies. Specifically, it contributes to the broad band of intensity between $\approx\!4$~meV and 6~meV,
which is in a close accord with the experimental observations of Refs.~\cite{nagler16,banerjee18} that interpreted it 
as a sign of fractionalization. The two-magnon continuum also  extends the observable bandwidth to the values 
that are consistent with experiments \cite{nagler16,Banerjee17,kaib19,Wulferding19}. 

The well-defined
magnon excitations at low energies are in a qualitative agreement with the neutron scattering results of 
 Refs.~\cite{wen17,banerjee18}. The weak $C_3$ symmetry breaking that is present in $\alpha$-RuCl$_3$
\cite{kaib19,winter16} is expected to provide a gap at the M points and suppress  the decays of the low-energy 
magnons near the $\Gamma$ point, thus improving the  agreement further.
The decays of the higher-energy magnons  and the broad band of intensity are expected to be insensitive 
to such a symmetry breaking. 

Altogether, our results strongly substantiate the expectations outlined in the beginning of this Section. 
The results of our calculations for a representative parameter set from the realistic parameter region 
yield the spectrum that is comprised of the broad excitation continua coexisting with the well-defined 
low-energy magnon modes. 
In accord with the scenario advocated in  Refs.~\cite{winter17, kopietz20}, 
it  highlights the significance  of the phenomenon of magnon decays and outlines
the challenges of interpreting broad features in the spectra of strongly-anisotropic magnets. 

%----------------------------------------------------------------------------
\begin{figure*}[t]
	 \includegraphics[width=.99\linewidth]{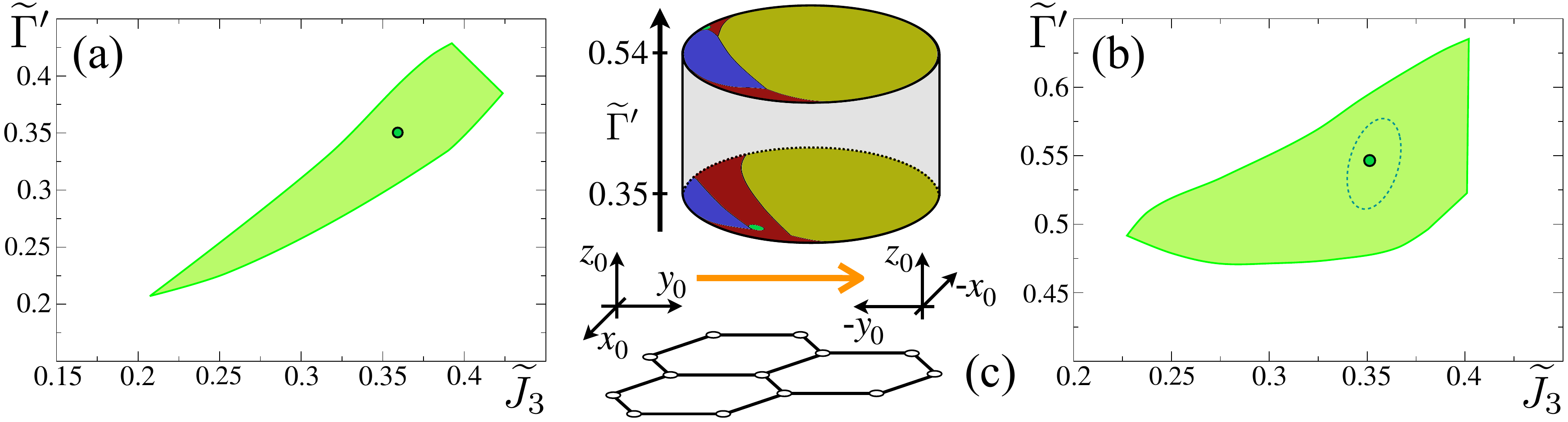} 
\vskip -0.2cm
\caption{(a) The $\widetilde{\Gamma}'$--$\widetilde{J}_3$ projection of the ``realistic'' ($W_0\!<\!15$~meV) parameter 
region  from  Sec.~\ref{Sec_model}, see also Figs.~\ref{fig_6}--\ref{fig_9}; 
$\widetilde{\Gamma}'$ and $\widetilde{J}_3$ are $\Gamma'$ and $J_3$  in units of  $\sqrt{J^2+K^2+\Gamma^2}$.
(b) Same  projection after  duality transformation of Eq.~(\ref{eq:pi_trans}). (c), 
bottom: a sketch of the $\pi$-rotation in the honeycomb plane about crystallographic $z_0$ axis.
(c), top: a schematic illustration of the transition between the two planes 
of the $J$--$K$--$\Gamma$ phase diagrams in Fig.~\ref{fig_duality2}
along the $\widetilde{\Gamma}'$-axis of an extended cylindrical 3D phase diagram as a result of the duality 
transformation.}
	\label{fig_duality1}
\vskip -0.4cm
\end{figure*}
%----------------------------------------------------------------------------

\vspace{-0.3cm}
% ==============================================================================
\section{Dual models and other forms}
\label{Sec_duality}
% ==============================================================================
\vskip -0.2cm

In this Section we provide  further insights into the relevant  section of the phase diagram 
of the generalized KH model that is pertinent to $\alpha$-RuCl$_3$ parameter space, analyze it with the help 
of the duality transformations, and discuss possible simplified versions of the effective model 
that should contain essential physics of this material. 

\vspace{-0.3cm}
% ==============================================================================
\subsection{Dualities}
% ==============================================================================
\vskip -0.2cm

The generalized KH model (\ref{H_JKGGp}) is known to map onto itself under various transformations 
\cite{Kimchi14, CJK13,ChKh15}, contributing to the aura of sophistication surrounding this model. 
While some of such transformations are rather non-trivial, others occur under benign symmetry operations,
with the latter made not obvious by %an unfortunate choice of  
the parametrization of the exchange 
matrix  \eqref{H_JKGGp}  in the cubic axes.

One of such artifacts of the cubic axes representation is the self-duality under the $\pi$-rotation of the honeycomb plane 
about crystallographic $z_0$ axis ([111] direction in the cubic axes), see Fig.~\ref{fig_axes} and  
Fig.~\ref{fig_duality1}. While for the model in crystallographic axes (see below) this innocent symmetry operation 
leads to an inconsequential sign change in one of the terms (${\sf J_{z\pm}}$), 
it requires rewriting of the generalized KH model \eqref{H_JKGGp} 
within the rotated set of the cubic axes  \cite{ChKh15}, transforming its parameters as%according to
\begin{eqnarray}
\label{eq:pi_trans}
\left( \begin{array}{c} J \\  K \\  \Gamma \\  \Gamma' \end{array} \right)_{\rm dual}=  \left( 
\begin{array}{cccc} 
1 & +\frac{4}{9} & -\frac{4}{9} & +\frac{4}{9} \\
0 & -\frac{1}{3} & +\frac{4}{3} & -\frac{4}{3} \\
0 & +\frac{4}{9} & +\frac{5}{9} & +\frac{4}{9} \\
0 & -\frac{2}{9} & +\frac{2}{9} & +\frac{7}{9} 
 \end{array} \right) \left( \begin{array}{c} J \\  K \\  \Gamma \\  \Gamma' \end{array} \right).
\end{eqnarray}
It is important to note that  models with the dual and  original parameters 
lead to  identical physical outcomes.

Because of that, the utility of such duality transformations is that they can reveal the origin of some 
properties of the generalized KH model that are hidden in the original language.  For instance, for the same model 
on the triangular lattice, the so-called Klein duality has allowed to relate an enigmatic
quasi-Goldstone  mode in the stripe phase  to an accidental degeneracy in the Klein-dual ferromagnetic
phase \cite{Zhu19}. 
In the spectrum of $\alpha$-RuCl$_3$, 
quasi-Goldstone modes are ubiquitously present at the M points that are complementary to the ordering vector of 
the zigzag phase \cite{wen17,banerjee18}, suggesting a proximity to  an accidental degeneracy.
This is also true throughout the  parameter space advocated in Sec.~\ref{Sec_model}, as is
highlighted in Fig.~\ref{fig_lswt3} for representative parameter sets.

Here, we use  duality transformations to shed light on the relevant phase diagram and properties
of $\alpha$-RuCl$_3$. In Fig.~\ref{fig_duality1}(a), we show 
$\widetilde{\Gamma}'$--$\widetilde{J}_3$ projection of the ``realistic''  parameter region, 
see Sec.~\ref{Sec_model}, where $\widetilde{\Gamma}'$ and $\widetilde{J}_3$ are 
$\Gamma'$ and $J_3$  normalized by $\sqrt{J^2+K^2+\Gamma^2}$, 
which is used as an energy scale, reducing the parameter space 
dimensions to 4D. Fig.~\ref{fig_duality1}(b) shows the same projection 
after duality transformation of Eq.~(\ref{eq:pi_trans}). 

\begin{figure}[t]
	 \includegraphics[width=.75\linewidth]{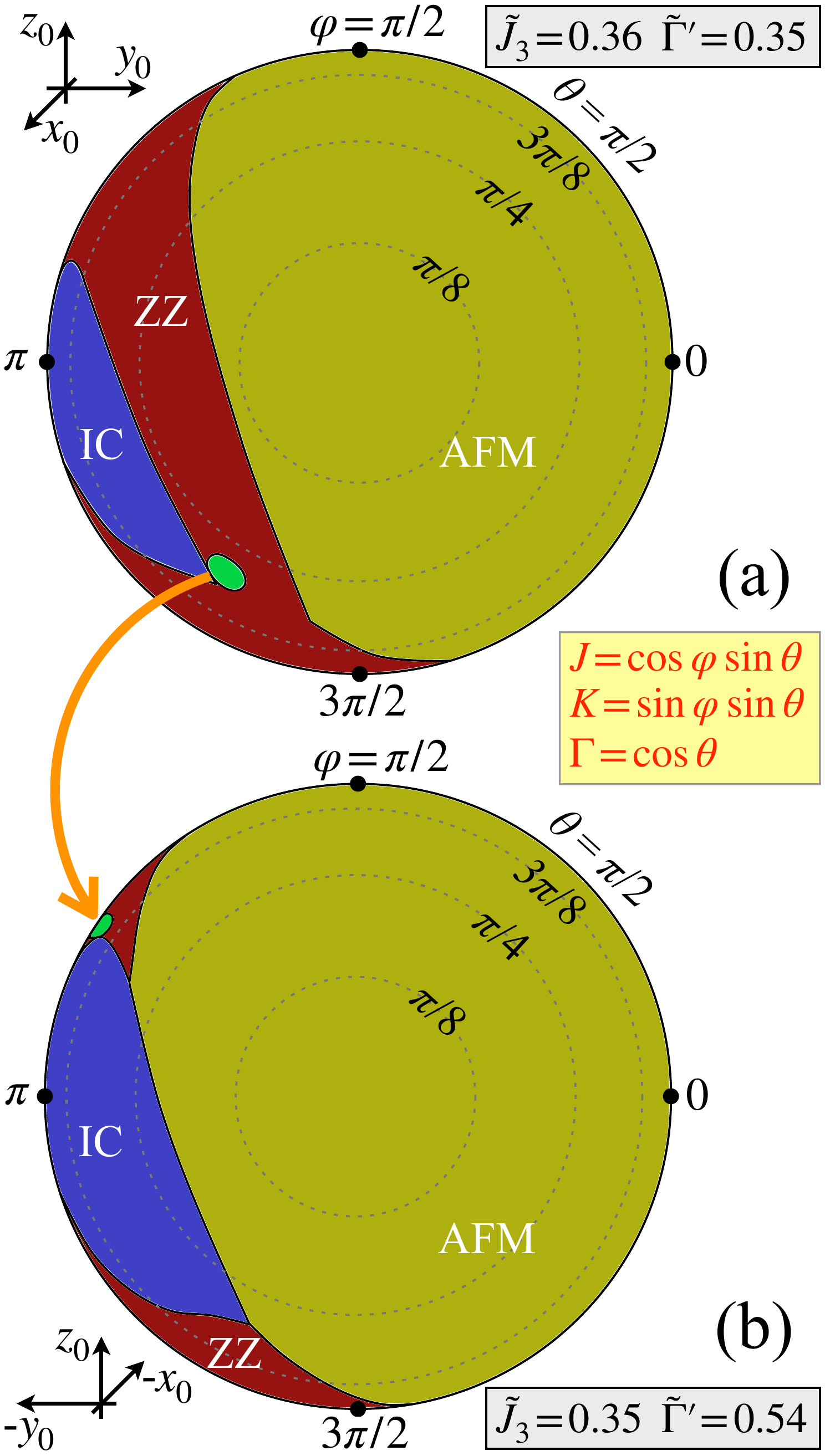} 
%\vskip -0.2cm
\caption{(a) The $J$--$K$--$\Gamma$ phase diagram for  
$\{\widetilde{\Gamma}',\widetilde{J}_3\}\!=\!\{0.35,0.36\}$, and (b) for 
$\{\widetilde{\Gamma}',\widetilde{J}_3\}\!=\!\{0.54,0.35\}$, see text.}
	\label{fig_duality2}
\vskip -0.2cm
\end{figure}

To make possible an exploration of the wider phase diagram, 
 we choose a representative point  from the ``original'' projection in Fig.~\ref{fig_duality1}(a), 
 $\{\widetilde{\Gamma}',\widetilde{J}_3\}\!=\!\{0.35,0.36\}$.  Since this fixes two parameters of the  4D parameter space, 
we can investigate the remaining  $J$--$K$--$\Gamma$ phase diagram by the Luttinger-Tisza method
\cite{LT}, see  Fig.~\ref{fig_duality2}(a) for the polar representation of it, in which
$\Gamma$ is the radial  and $J$ and $K$ are the polar variables. 

The entire phase space is exhausted by the aniferromagnetic (AFM), zigzag (ZZ), and the incommensurate (IC) states.
The parameter space associated with the ``realistic'' parameter choices occupies a small region 
of the zigzag phase bordering    incommensurate phase, as is discussed 
in Sec.~\ref{Sec_model}~D, see also Appendix~\ref{app_0}. 
The zigzag-to-incommensurate phase transition is of the first order by both LT and LSWT analysis. The IC 
phase evolves  continuously from a ferromagnetic state in a broader $\widetilde{\Gamma}'$--$\widetilde{J}_3$
parameter space and is similar in nature to the  helical phase within 
the phase diagram of the $J_1$--$J_2$--$J_3$ model on the same lattice \cite{italians79}.

It is the dual $\pi$-rotated version of this phase diagram that is of interest. A minor subtlety occurs because the 
transformation of Eq.~(\ref{eq:pi_trans}) concerns all four parameters of the exchange matrix, so that the 
transformation of the single point in  Fig.~\ref{fig_duality1}(a) results in an area 
in the dual $\widetilde{\Gamma}'$--$\widetilde{J}_3$ projection in Fig.~\ref{fig_duality1}(b), which is highlighted by the ellipse.
To make a comparison meaningful and given a small size of the dual 
region of interest, we also pick a representative point in the dual region of parameters,
$\{\widetilde{\Gamma}',\widetilde{J}_3\}\!=\!\{0.54,0.35\}$, see Fig.~\ref{fig_duality1}(b). 
We also note that since the duality transformation (\ref{eq:pi_trans}) involves four parameters, one can see it as a  
transition between the planes of a 3D cylindrical phase diagram along the $\widetilde{\Gamma}'$-axis, 
as is schematically illustrated in Fig.~\ref{fig_duality1}(c).

%----------------------------------------------------------------------------
%----------------------------------------------------------------------------

The dual $J$--$K$--$\Gamma$ phase diagram for this choice of $\widetilde{\Gamma}'$ and $\widetilde{J}_3$ 
is shown in  Fig.~\ref{fig_duality2}(b).  The most important observation  
is that the duality-transformed ``realistic'' parameter region of the generalized KH 
model  corresponds to the vanishing values of the $\Gamma$-term. 
This  is also corroborated by the dual transformation (\ref{eq:pi_trans}) 
of the Point 1, 2, and 3 representative parameter sets of the 
model from Sec.~\ref{Sec_model}~E, given by the  following dual
$(K,\Gamma,\Gamma',J,J_3)$ coordinates (in meV)
\vskip -0.2cm
\begin{center}
\vskip -0.2cm
\begin{tabular}{lrrrrr}
      & $(K,$  & $\Gamma,$   & $\Gamma',$  &  $J,$  & $J_3)\ $   \\ 
{dual Point 1:} & (3.71, & 1.24, & 3.92, & -5.40, & 2.42) \  \\ 
{dual Point 2:} & (6.67, & -0.62, & 5.81, & -9.82, & 3.26) \  \\ 
{dual Point 3:} & (8.72, & -1.72, & 7.20, & -12.32, & 3.66).   
\end{tabular}
\end{center}
\vskip -0.2cm
In all three cases, as well as in Fig.~\ref{fig_duality2}(b), the allowed parameters  correspond to 
$\Gamma\!\approx\!0$, dominant 
\emph{ferromagnetic} $J\!<\!0$, followed by   substantial \emph{antiferromagnetic} Kitaev 
and  $\Gamma'$ terms, $K\!\approx\!\Gamma'\!>\!0$.  
The dual zigzag region is also bordered by the incommensurate 
phase because the global $\pi$-rotation does not affect the ground state. 
We emphasize once more that the resultant physical properties of the 
original and dual models are identical. 

The duality of the original model  to the one with negligible $\Gamma$ 
sheds new light onto the nature of the persistent pseudo-Goldstone modes in $\alpha$-RuCl$_3$
\cite{wen17,banerjee18}. The pure 
Kitaev-Heisenberg model with $\Gamma\!=\!\Gamma'\!=\!0$ is well-known to have a classical 
accidental degeneracy leading to a gapless mode \cite{Kimchi14, CJK13}.
Such a degeneracy is generally lifted by $\Gamma\!\neq\!0$. It is less known that the non-zero
$\Gamma'$ term can leave this degeneracy intact \cite{RauGp}.  
The logic of this behavior is exposed by the four-sublattice  Klein duality transformation 
that converts  zigzag to  antiferromagnetic state \cite{Kimchi14, CJK13}. 
While this transformation maps pure Kitaev-Heisenberg model onto itself, 
the off-diagonal $\Gamma'$ term morphs into antisymmetric, 
Dzyaloshinskii-Moriya-like interaction  $\breve{\Gamma}'$. The latter  does not affect collinear ground state on 
a classical level, leaving the accidental degeneracy intact and preserving pseudo-Goldstone modes.
This effect is  similar to the one discussed for the same model on the triangular lattice \cite{Zhu19}. 

%\vspace{-0.3cm}
% ==============================================================================
\subsection{Simpler models}
% ==============================================================================
%\vskip -0.2cm

The other  reason why the negligible-$\Gamma$ form of the generalized KH model is important, is because 
it results in an effective description of $\alpha$-RuCl$_3$ by fewer parameters. The relevant 
$K$--$J$--$\Gamma'$--$J_3$ model  has a lower dimensionality of its parameter space and 
is, thus, more amendable to a detailed exploration.

A potentially more drastic simplification  can be achieved by rewriting 
the   model (\ref{eq_Hij}) 
in the ``spin-ice'' language that uses crystallographic axes
\cite{Ross11,RauGp,ChKh15,RauYb,Tsirlin_Review,Zhu19}, where $x_0$ and $y_0$ correspond to the 
$a$ and $b$ directions in the plane of the honeycomb lattice, see Fig.~\ref{fig_axes}.
This leads to the $XXZ$--${\sf J_{\pm\pm}}$--${\sf J_{z\pm}}$ form of the  Hamiltonian \eqref{eq_Hij} 
\begin{align}
\label{HJpm}
&{\cal H}_1=\sum_{\langle ij\rangle} \Big[
{\sf J_1} \Big(S^{x_0}_i S^{x_0}_j+S^{y_0}_i S^{y_0}_j+\Delta S^{z_0}_i S^{z_0}_j\Big)\\
&\!-\!2 {\sf J_{\pm \pm}} 
\Big( \Big( S^{x_0}_i S^{x_0}_j \!- \!S^{y_0}_i S^{y_0}_j \Big) \tilde{c}_\alpha 
\!-\!\Big( S^{x_0}_i S^{y_0}_j\!+\!S^{y_0}_i S^{x_0}_j\Big)\tilde{s}_\alpha \Big)\nonumber\\
&\!-\!{\sf J_{z\pm}}\Big( \Big( S^{x_0}_i S^{z_0}_j \!+\!S^{z_0}_i S^{x_0}_j \Big) \tilde{c}_\alpha 
 \!+\!\Big( S^{y_0}_i S^{z_0}_j\!+\!S^{z_0}_i S^{y_0}_j\Big)\tilde{s}_\alpha \Big)\Big]\nonumber,
\end{align}
where abbreviations are $\tilde{c}_\alpha\!=\!\cos\tilde{\varphi}_\alpha$ and 
$\tilde{s}_\alpha\!=\!\sin\tilde{\varphi}_\alpha$, bond-dependent phases 
$\tilde{\varphi}_\alpha\!=\!\{0,2\pi/3,-2\pi/3\}$  correspond to the $\{{\rm Z,X,Y}\}$ bonds  in Fig.~\ref{fig_axes}, 
respectively, and the isotropic third-neighbor ${\cal H}_3$ is unchanged.

The relation of the  parameters of the model (\ref{HJpm})  to the parameters of the model (\ref{H_JKGGp})
is given in Appendix~\ref{app_A}. 
Rewriting the Point 1, 2, and 3 representative parameter sets from Sec.~\ref{Sec_model}~E in these new variables yields  
\vskip -0.2cm
\begin{center}
\vskip -0.4cm
\begin{tabular}{lrrrrr}
      & $({\sf J_1},$  & $\Delta,$   & ${\sf J_{\pm \pm}},$  &  ${\sf J_{z\pm}},$  & $J_3)\ $   \\ 
{``ice'' Point 1:} & (-7.20, & -0.26, & 0.3, & -3.0, & 2.42) \  \\ 
{``ice'' Point 2:} & (-11.3, & 0.02, & 1.0, & -6.2, & 3.26) \  \\ 
{``ice'' Point 3:} & (-13.6, & 0.07, & 1.5, & -8.3, & 3.66),   
\end{tabular}
\end{center}
\vskip -0.2cm
all in meV except for the dimensionless $XXZ$ anisotropy parameter $\Delta$.

It transpires that for all three representative sets,  the easy-plane anisotropy $\Delta$ is  small 
and one of the bond-dependent terms ${\sf J_{\pm\pm}}$ is much smaller than the other interactions.
We can verify that the same is true across the advocated realistic parameter ranges for 
$\alpha$-RuCl$_3$: the model  in the language of Eq.~(\ref{HJpm}) consistently has these two terms
nearly negligible. Some of the parameter sets suggested in the prior works based on $\alpha$-RuCl$_3$
phenomenology also follow the same trend \cite{winter17}.
Of the remaining terms, the leading is the ferromagnetic $XY$ exchange ${\sf J_1}$ 
with the sizable ${\sf J_{z\pm}}\!\approx\!|{\sf J_1}|/2$ and $J_3$.

It follows from this analysis that the ${\sf J_1}$--${\sf J_{z\pm}}$--$J_3$ model, 
written in   crystallographic axes of the honeycomb lattice and
operating in a much more accessible 3D parameter space, should be able to 
offer a  much simpler and much more natural way of describing $\alpha$-RuCl$_3$ 
that can give a refreshing perspective on its physics. 

It also appears that the relevant physics of $\alpha$-RuCl$_3$ is not related to a model with a dominating 
Kitaev term, but is that of the easy-plane ferromagnet with antiferromagnetic third-neighbor coupling 
and strong off-diagonal  exchange ${\sf J_{z\pm}}$. Naturally, it is the latter term that is responsible 
for the out-of-plane tilting of the ordered moment, substantial  fluctuations 
in the ground state, and significantly damped magnon excitations. 

From this study,  one is also led to believe that it is not a proximity to a Kitaev spin-liquid phase,
but a proximity to an incommensurate phase, which is continuously connected to a ferromagnetic one,  
that is significantly more pertinent to the phase diagram of $\alpha$-RuCl$_3$.
These and other features of this material and relevant models deserve further investigation.

%\vspace{-0.3cm}
% ==============================================================================
\section{Summary}
\label{Sec_conclusions}
% ==============================================================================
%\vskip -0.2cm

We conclude by summarizing our results. 

We have demonstrated that empirical constraints lead to 
significant restrictions and rather drastic revisions of the physically reasonable parameter space for the 
effective microscopic spin model of  $\alpha$-RuCl$_3$. 
Specifically, the ESR and THz data in the field-induced paramagnetic regime,
combined with the analysis of the in-plane critical fields, out-of-plane tilt angle, bandwidth of the magnetic signal,
and the zigzag nature of the ground state, produce convincing bounds on the parameters of this model.

In broad strokes, for the key parameters of the generalized KH model,
these constraints necessitate a significant positive $\Gamma'\!\approx\!\Gamma/2$ 
not anticipated previously and a close relation $J\!+\!3J_3\!\approx\!\Gamma$.  
The leading Kitaev term $K\!<\!0$ is also constrained and is not overly dominating, 
with the ratio $|K|/\Gamma\!\approx\! 1.0-3.0$, and $J\!<\!0$ and $J_3\!>\!0$ terms varying 
within the range of $|J|/\Gamma\! \approx\! J_3/\Gamma\! \approx\!0.5-0.75$.
In the absolute units, the bounds on $\Gamma$ are $\approx\!4$~meV$-$6~meV, setting the scale for the
rest of the model.

We have also demonstrated that our proposed parameter sets provide an excellent account of a variety 
of other phenomenologies of $\alpha$-RuCl$_3$, allowing to consolidate previous attempts of their description.
Our parameters can  be reconciled with the typically smaller
parameters discussed in the prior works by suggesting their renormalization  due to quantum fluctuations. 
We have shown that the latter effect can be successfully approximated with the help of a self-consistent 
mean-field RPA approach.

We have argued, in accord with the previous studies, that the off-diagonal terms that 
necessarily produce strong anharmonic couplings of magnons are   expected to be significant
throughout the phase diagram of an anisotropic-exchange magnet and for any form of the underlying magnetic order. 
These couplings, in turn, inevitably lead to large decay rates of the higher-energy   
into the lower-energy magnons, resulting in a coexistence of the 
broad continua with the well-defined low-energy  modes in the inelastic neutron scattering 
spectrum. The proposed parameter space of $\alpha$-RuCl$_3$ is no exception to this scenario, 
with our calculations of the dynamical structure factor for a representative parameter set strongly substantiating it 
in a close agreement with experiments. This result highlights the challenges of interpreting broad features in 
the spectra of strongly-anisotropic magnets and the significance of the phenomenon of 
magnon decays in this context.

We have also provided an important insight into the nature of the pseudo-Goldstone modes that occur 
away from the ordering vector of the zigzag phase of $\alpha$-RuCl$_3$.
Using  duality transformations of the generalized KH model within the advocated parameter range, 
we have related these modes to an accidental near-degeneracy in a duality-related $\Gamma$-less model, 
in which the degeneracy of the pure Kitaev-Heisenberg type is not lifted by the $\Gamma'$ term.
As a by-product, this effort has suggested  a fully equivalent simpler model description 
of $\alpha$-RuCl$_3$ within the same KH model, but with the leading term $J\!<\!0$, 
subleading positive $K\!\approx\!\Gamma'$, finite $J_3$, and negligible $\Gamma$. 

A different and substantially more radical simplification advocated in this work is the rewriting of the 
generalized KH model in the natural crystallographic axes of the honeycomb lattice.
We have verified that for the advocated realistic parameter ranges of 
$\alpha$-RuCl$_3$, the model  in this language has only three substantial terms: 
the leading ferromagnetic easy-plane ${\sf J_1}$, antiferromagnetic  $J_3$, 
and a sizable off-diagonal term ${\sf J_{z\pm}}$. The latter term favors 
the observed out-of-plane tilting of spins in the zigzag phase and its role 
in strong quantum effects and magnon interactions deserves further investigation.

Thus, one of the key result of the rethinking endeavor  undertaken in this work
is that the relevant physics of $\alpha$-RuCl$_3$ is not related to a model with a dominating 
Kitaev term, but must be understood and revisited as that of the ${\sf J_1}$--$J_3$ FM-AFM model with
the dominant easy-plane ${\sf J_1}$ and a strong off-diagonal  exchange ${\sf J_{z\pm}}/|{\sf J_1}|\!\approx\!0.5$. 

Altogether, the provided consideration of the $\alpha$-RuCl$_3$ phenomenologies and its effective model
unequivocally suggests that the physics of this material is not affiliated with a proximate spin-liquid state.
The only proximity in the phase diagram that is present in this case and may be worth exploring   
is that to an incommensurate phase, which is continuously connected to a ferromagnetic one.
For the much-discussed spectral properties of $\alpha$-RuCl$_3$, the conclusion of this work also unambiguously 
points toward the physics of the strongly interacting and mutually decaying magnons, not to that of the 
fractionalized excitations.

%\vspace{-0.3cm}

% ==============================================================================
\begin{acknowledgments}
% ==============================================================================
%\vskip -0.2cm

We would like to thank Sergei Zvyagin and Paul van Loosdrecht for sharing their 
experimental data that were instrumental for our proposed parameter constraints.
We are indebted to Stephen Winter, David Kaib, and Roser Valent{\'{\i}}  for numerous fruitful discussions,
for sharing their ED data to benchmark our results, for providing an important first-principles support 
of our  large $\Gamma'\!>\!0$ argument, and also for an immense patience with the pace of 
our progress on this work. 
We are especially grateful to Jeff Rau for reminding us about the value of dualities that has resulted 
in Sec.~\ref{Sec_duality} and for lending us a generous numerical  and psychological support on the 
validity of our quantum renormalization arguments that is not included in this study. We 
would also like to thank Qiang Luo for spotting several inconsequential typos in Appendix~\ref{app_A} 
and for communicating them in a peaceful manner.

This work was supported by the U.S. Department of Energy,
Office of Science, Basic Energy Sciences under Award No. DE-FG02-04ER46174 (P. A. M. and A. L. C.).
P.~A.~M. acknowledges support from JINR Grant for young scientists 20-302-03.
A.~L.~C. would like to thank Aspen Center for Physics and the Kavli Institute for Theoretical Physics (KITP)
where different stages of this work were advanced. 
The Aspen Center for Physics is supported by National Science Foundation Grant No. PHY-1607611 
and KITP is supported  by the National Science Foundation under  Grant No. NSF PHY-1748958.

% ==============================================================================
\end{acknowledgments}
% ==============================================================================

\appendix

%\vspace{-0.3cm}

% ==============================================================================
\section{LSWT details}
\label{app_A}
% ==============================================================================
%\vskip -0.2cm

The ``spin-ice'' form of the Hamiltonian \eqref{H_JKGGp} in the crystallographic $\{x_0,y_0,z_0\}$ axes 
of the honeycomb plane is given by Eq.~(\ref{HJpm}). Its parameters are related to that of the generalized KH model in 
the cubic axes \eqref{H_JKGGp} via
\begin{align}
{\sf J_1}&=J+\frac{1}{3} \big( K-\Gamma-2\Gamma'\big),\nonumber\\
\Delta {\sf J_1}&=J+\frac{1}{3} \big( K+2\Gamma+4\Gamma'\big),\nonumber\\
\label{eq_JpmKG}
2{\sf J_{\pm\pm}}&=-\frac{1}{3} \big( K+2\Gamma-2\Gamma'\big),\\
\sqrt{2}{\sf J_{z\pm}}&=\frac{2}{3} \big( K-\Gamma+\Gamma'\big).\nonumber
\end{align}
The rotation to the local reference frame of spins for the field-induced polarized paramagnetic state
with the subsequent Holstein-Primakoff and Fourier transformations in \eqref{HJpm} 
yield the LSWT Hamiltonian  
\begin{align}
\mathcal{H} =\frac{1}{2}\sum_\mathbf{k}\mathbf{x}_\mathbf{k}^\dagger 
\mathbf{H_k}\mathbf{x}_\mathbf{k}^{\phantom{\dagger}},
\label{eq_hamak}
\end{align}
where $\mathbf{x}_\mathbf{k}\!=\!\big( a^{\phantom{\dagger}}_\mathbf{k},
b^{\phantom{\dagger}}_\mathbf{k},a^\dagger_\mathbf{-k},b^\dagger_\mathbf{-k}\big)$, and $a_\mathbf{k}$ and 
$b_\mathbf{k}$ are bosonic magnon operators on two sublattices of the honeycomb lattice. 
The   form of the Hamiltonian   in the polarized phase for the principal in-plane field directions is  
\begin{eqnarray}
\mathbf{H}=
\left( \begin{array}{cccc} 
A_\mathbf{k} &  B_\mathbf{k} & 0 & C_\mathbf{k}\\ 
B^*_\mathbf{k} &  A_\mathbf{k} & C_\mathbf{-k} & 0\\
0 &  C^*_\mathbf{-k} & A_\mathbf{k} & B_\mathbf{k}\\
C^*_\mathbf{k} &  0 & B^*_\mathbf{k} & A_\mathbf{k}
\end{array}\right),
\label{eq_hammatrix}
\end{eqnarray}
where for $H \!\parallel\! a$
\begin{align}
A_\mathbf{k}&=g\mu_B H-3S\big({\sf J_1}+J_3\big),\nonumber\\
\label{eq_Ha}
B_\mathbf{k}&=\frac{3S}{2} \left[ {\sf J_1}\big(1+\Delta\big) \gamma_\mathbf{k} 
+2 J_3 \gamma^{(3)}_\mathbf{k}+2{\sf J_{\pm\pm}}\gamma'_\mathbf{k}\right],\\
C_\mathbf{k}&=-\frac{3S}{2} \left[ {\sf J_1}\big(1-\Delta\big) \gamma_\mathbf{k} 
+2{\sf J_{\pm\pm}}\gamma'_\mathbf{k}-2i{\sf J_{z\pm}}\gamma''_\mathbf{k}\right],\nonumber
\end{align}
and for $H \!\parallel \!b$
\begin{align}
A_\mathbf{k}&=g\mu_B H-3S\big({\sf J_1}+J_3\big),\nonumber\\
\label{eq_Hb}
B_\mathbf{k}&=\frac{3S}{2} \big[ {\sf J_1}\big(1+\Delta\big) \gamma_\mathbf{k} 
+2J_3 \gamma^{(3)}_\mathbf{k}-2{\sf J_{\pm\pm}}\gamma'_\mathbf{k}\big],\\
C_\mathbf{k}&=\frac{3S}{2} \big[ {\sf J_1}\big(1-\Delta\big)\gamma_\mathbf{k} 
-2{\sf J_{\pm\pm}}\gamma'_\mathbf{k}+2i{\sf J_{z\pm}}\gamma'_\mathbf{k}\big],\nonumber
\end{align}
where the hopping functions are given by
\begin{align}
\gamma_\mathbf{k}&=\frac{1}{3}\sum_{\alpha=1}^3 e^{i\mathbf{k}{\bm \delta}_\alpha}, \ \ \
\gamma^{(3)}_{\mathbf{k}}=\frac{1}{3}\sum_{\alpha=1}^3 e^{i\mathbf{k}{\bm \delta}^{(3)}_\alpha},\\
\gamma'_\mathbf{k}&=\frac{1}{3}\sum_{\alpha=1}^3 \cos\tilde{\varphi}_\alpha e^{i\mathbf{k}{\bm \delta}_\alpha}, \ \ \
\gamma''_\mathbf{k}=\frac{1}{3}\sum_{\alpha=1}^3 \sin\tilde{\varphi}_\alpha e^{i\mathbf{k}{\bm \delta}_\alpha},
\end{align}
and ${\bm \delta}_{\alpha}$  are the vectors connecting nearest-neighbor sites along the 
${\{\rm Z,X,Y\}}$ bonds,  respectively, see Fig.~\ref{fig_axes}.

The LSWT spectrum is given by the standard procedure for a bosonic Hamiltonian \cite{Colpa}, which requires   
diagonalization of $\mathbf{g}\mathbf{H_k}$, where $\mathbf{g}$ is a diagonal matrix $\{1,1,-1,-1\}$.  
For $\mathbf{H_k}$ in Eq.~(\ref{eq_hammatrix}) this diagonalization can be done analytically 
and the eigenvalues are given by the solutions of the biquadratic equation
\begin{align}
\lambda^4-2k\lambda^2+c=0,
\end{align}
where 
\begin{align}
k&=A_\mathbf{k}^2+\big| B_\mathbf{k} \big|^2-\frac{\big| C_\mathbf{k} \big|^2+\big| C_\mathbf{-k} \big|^2}{2},\\
c&=\big( A_\mathbf{k}^2-\big| B_\mathbf{k} \big|^2\big)^2-
A_\mathbf{k}^2\big( \big| C_\mathbf{k} \big|^2+\big| C_\mathbf{-k} \big|^2 \big)\nonumber\\
&+\big| C_\mathbf{k} \big|^2 \big| C_\mathbf{-k} \big|^2
-B_\mathbf{k}^2 C_\mathbf{-k} C^*_\mathbf{k}-(B^*_\mathbf{k})^2 C^*_\mathbf{-k} C_\mathbf{k}.\nonumber
\end{align}
These expressions simplify for the high-symmetry $\mathbf{k}$ points. 
First,  the magnon spectrum at the $\Gamma$ point, $\mathbf{k}\!=\!0$, for both field directions is 
\begin{align}
\varepsilon_{1,2}=\sqrt{\big( A_{\mathbf{k} }\pm B_{\mathbf{k} }\big)^2-\big|C_{\mathbf{k} } \big|^2}.
\label{eq_e_gamma}
\end{align}
In particular, the lowest mode as a function of magnetic field is given by
\begin{align}
\varepsilon_{1,\mathbf{k}=0}&=\sqrt{g\mu_B H \big( g\mu_B H-3{\sf J_1}S(1-\Delta)\big)}.
\end{align}
Translation to the generalized KH interactions via \eqref{eq_JpmKG} yields Eq.~\eqref{Ek0} in Sec.~\ref{Sec_constraints}.

For the field $H\!\parallel\! a$, the 
important high-symmetry points are the $M$ points, $\mathbf{k}_{M(M')}\!=\!\big(\pi/\sqrt{3},\pm \pi \big)$. 
The matrix elements in Eq.~(\ref{eq_Ha}) for these points, with 
the help of the transformation $b_\mathbf{k}\!\rightarrow\! b_\mathbf{k} e^{\pm i\pi/3}$, simplify to  
\begin{align}
A_\mathbf{k}&=g\mu_B H-3S\big({\sf J_1}+J_3\big),\nonumber\\
\label{eq_Ha1}
B_\mathbf{k}&=\frac{S}{2}\big[ {\sf J_1}\big(1+\Delta\big) -6J_3 +2 {\sf J_{\pm\pm}}\big],\\
C_\mathbf{k}&=-\frac{S}{2} \big[ {\sf J_1}\big(1-\Delta\big)  
+2 {\sf J_{\pm\pm}}+ 2\sqrt{3} i{\sf J_{z\pm}}\big],\nonumber
\end{align}
and the magnon energies are given by the same Eq.~\eqref{eq_e_gamma}. 
From Eqs.~\eqref{eq_e_gamma} and (\ref{eq_Ha1}),  one obtains a transition field from the polarized  
to  zigzag phase by finding $H_c$ that corresponds to closing of the gap at the $M$ points, 
$\varepsilon_{1,\mathbf{k}}\!=\!0$,  resulting in Eq. \eqref{Hca}.

For the  field in the $b$-direction, the gap closes at the $Y$ point and the matrix elements for 
$\mathbf{k}_{Y}\!=\!\big(2\pi/\sqrt{3},0 \big)$ in Eq.~(\ref{eq_Hb})  are given by
\begin{align}
A_\mathbf{k}&=g\mu_B H-3S({\sf J_1}+J_3),\nonumber\\
B_\mathbf{k}&=\frac{S}{2}\big[ {\sf J_1}\big(1+\Delta\big) -6J_3 +4 {\sf J_{\pm\pm}}\big],\\
C_\mathbf{k}&=\frac{S}{2} \big[ {\sf J_1}\big(1-\Delta\big)  +4 {\sf J_{\pm\pm}}-4 i{\sf J_{z\pm}}\big].\nonumber
\end{align}
The solution for the gap closure gives the critical field in the $b$-direction, Eq.~\eqref{Hcb}.

% ==============================================================================
\section{Self-consistent calculation of the N\'eel  temperature}
\label{app_TN}
% ==============================================================================

The spin Green's function in the self-consistent RPA approximation  is given by \cite{Tyablikov,Plakida,Zhu19}
\begin{align}
\label{eq_gf}
\langle S^{-}_\mathbf{k} S^+_\mathbf{-k} \rangle_\omega=2\langle S \rangle\sum_{\mu} 
\left(\frac{u_{\mu\mathbf{k}}^2}{\omega - \widetilde{\varepsilon}_{\mu\mathbf{k}}}-
\frac{v_{\mu\mathbf{k}}^2}{\omega +\widetilde{\varepsilon}_{\mu\mathbf{k}}}\right),
\end{align}
where $u_{\mu\mathbf{k}}$ and $v_{\mu\mathbf{k}}$ are parameters of the generalized 
Bogolyubov transformation that diagonalizes the spin-wave bosonic Hamiltonian, $\mu$ numerates
bosonic branches in the zigzag phase, 
and $\widetilde{\varepsilon}_{\mu\mathbf{k}}\!=\!2\langle S \rangle\varepsilon_{\mu\mathbf{k}}$, 
see Eq.~\eqref{eq_rpa}.

A self-consistent condition on the ordered moment is obtained from 
$S^z_i =\frac{1}{2}-S^{-}_i S^+_i$ 
using spectral representation in Eq.~\eqref{eq_gf}, resulting in 
\begin{eqnarray}
\label{eq_app_rpa_s}
\langle S \rangle =\frac{1}{2}-
\frac{2\langle S\rangle}{N}  \sum_{\mu,\mathbf{k}} 
\Big( u_{\mu\mathbf{k}}^2 n\left(\widetilde{\varepsilon}_{\mu\mathbf{k}} \right)-
v_{\mu\mathbf{k}}^2 n\left(-\widetilde{\varepsilon}_{\mu\mathbf{k}} \right) \Big). \ \ \ \ \ \ 
\end{eqnarray}
where $n(\omega)$ is the Bose distribution function. 

At $T\!\rightarrow\!T_N$, the ordered moment $\langle S \rangle\!\rightarrow\! 0$ 
and (\ref{eq_app_rpa_s}) yields the ordering  temperature 
\begin{align}
\frac{1}{T_N}=\frac{2}{N}\sum_{\mu,\mathbf{k}} \frac{u_{\mu\mathbf{k}}^2+
v_{\mu\mathbf{k}}^2}{\varepsilon_{\mu\mathbf{k}}}.
\label{eq_tn}
\end{align}
Note that the outlined approach is  valid only for  $S\!=\!1/2$, with the formalism becoming more complicated 
for larger spins \cite{Tyablikov}.

%\vspace{-0.3cm}
% ==============================================================================
\section{$J$--$J_3$ phase diagram for fixed $\{K,\Gamma,\Gamma'\}$}
\label{app_0}
% ==============================================================================
%\vskip -0.2cm

A useful insight can be provided by using a ``strict'' form of the proposed constraints and examining the 
remaining low(er)-dimensional parameter space for its phase diagram. 
This approach can also help to alleviate a concern that the 2D projections from a higher-dimensional parameter space 
with the dimensions higher than 3D can give a false perception of the 
phase diagram. This is because such 2D projections simply demonstrate the largest possible extent
of the allowed parameters, which can be significantly different for different lower-dimensional 
``cuts'' of the higher-dimensional object.  
 
Here, we use our constraints on ESR/THz gap, $\Delta H_c$, and tilt-angle $\alpha$, by strictly 
fixing them to the values that are close to, or motivated by, the experiments. 
Thus, we fix $\Gamma_{\rm tot}\!=\!9$~meV, $\Delta H_c\!=\!1$~T, and $\alpha\!=\!35^{\degree}$
according to the discussions in Secs.~\ref{Sec_model}~A--D.
Since, quasiclassically, these quantities depend only on $K$, $\Gamma$, and $\Gamma'$ combinations, this specific 
choice of constraints yields the following set of $\{K,\Gamma,\Gamma'\}\!=\!\{-7.567, 4.276, 2.362\}$~meV.
We will refer to it as to the ``Point 0 set'' below.
  
%----------------------------------------------------------------------------
\begin{figure}[t]
	 \includegraphics[width=1\linewidth]{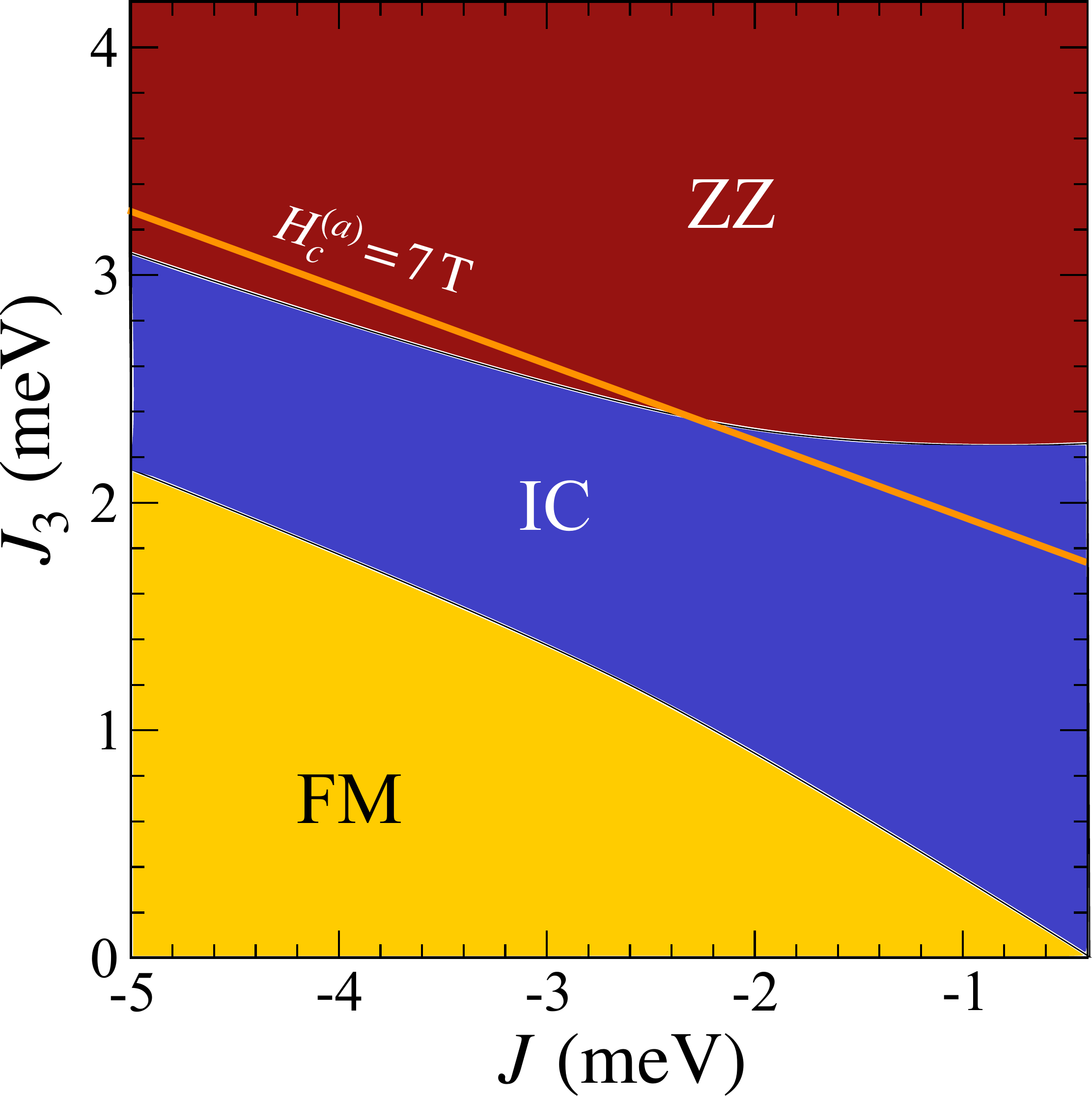} 
\vskip -0.2cm
\caption{The $J$--$J_3$ phase diagram for  $J\!<\!0$ and $J_3\!>\!0$
at fixed $\{K,\Gamma,\Gamma'\}\!=\!\{-7.567, 4.276, 2.362\}$~meV, see text.
The highlighted regions are the zigzag (ZZ), ferromagnetic (FM), and 
incommensurate phases (IC).
}
	\label{fig_J_J3}
\vskip -0.4cm
\end{figure}
%----------------------------------------------------------------------------

Of the 5D parameter space,  only $J$ and $J_3$ parameters remain. This allows us to explore the 
2D $J$--$J_3$ phase diagram for the relevant ranges of $J\!<\!0$ and $J_3\!>\!0$, presented in Fig.~\ref{fig_J_J3}.
The highlighted regions correspond to the zigzag (ZZ), ferromagnetic (FM), and %various forms of 
incommensurate (IC) phases.
Our last constraint that can be made rigid is to  fix 
$H_c^{(a)}\!=\!H_c^{(a)}(K,\Gamma,\Gamma',J_{03})$ to 
its experimental value of  $H_{c,{\rm exp}}^{(a)}\!=\!7$T, see Sec.~\ref{Sec_model}~C.
This binds $J_{03}\!=\!J+3J_3$ from Eq.~(\ref{J03}), yielding  
$J_{03}\!=\!4.768$~meV and restricting $J$ and $J_3$ to the straight line shown in Fig.~\ref{fig_J_J3}.

Altogether, this consideration illustrates that the values of $|J|$ and $J_3$ that are needed 
to stabilize the zigzag phase are larger than is typically assumed, see Table~\ref{table1}. 
Another important observation is that the empirically-constrained parameter sets put   
$\alpha$-RuCl$_3$ in the proximity of an incommensurate phase. As is discussed in Sec.~\ref{Sec_duality},
this phase is reminiscent of that in the phase diagram of the $J_1$--$J_2$--$J_3$ model on the 
honeycomb lattice and is continuously connected to a ferromagnetic state.

\vspace{-0.3cm}

% ==============================================================================
\section{Off-diagonal terms in Eq.~(\ref{H3})}
\label{app_B}
% ==============================================================================
\vspace{-0.3cm}

There are five 
distinct bonds regarding the values of $\widetilde{J}_{ij}^{xz}$, $\widetilde{J}_{ij}^{yz}$ or their combinations.
Keeping explicit the angles $\varphi$ and $\theta$, which are defined as polar and azimuthal angles relative to 
cubic axes, the real-space three-magnon couplings in Eq.~(\ref{H3}) for the bonds ${\rm{X},\rm{Y},\rm{Z}}$ are given by
\begin{align}
\mbox{{\tt ab}{\rm{X}}}:\quad&\widetilde{J}_{ij}^{xz}=-(\Gamma\sin\varphi+\Gamma'\cos\varphi)\cos2\theta\nonumber\\
&\quad\quad\ \  +(K\cos^2\varphi+\Gamma' \sin 2\varphi)\sin\theta\cos\theta,\nonumber\\
   \,      &\widetilde{J}_{ij}^{yz}=(\Gamma\cos\varphi-\Gamma' \sin\varphi)\sin\theta\nonumber\\
   &\quad\quad\ \   -(K\sin\varphi\cos\varphi-\Gamma' \cos2\varphi)\cos\theta,\nonumber %\\
\end{align}
\begin{align}
\mbox{{\tt ab}{\rm{Y}}}:\quad&\widetilde{J}_{ij}^{xz}=-(\Gamma\cos\varphi+\Gamma'\sin\varphi)\cos2\theta\nonumber\\
&\quad\quad\ \   +(K\sin^2\varphi+\Gamma'\sin2\varphi)\sin\theta\cos\theta,\nonumber\\
   \,      &\widetilde{J}_{ij}^{yz}=-(\Gamma\sin\varphi-\Gamma'\cos\varphi)\sin\theta\nonumber\\
   &\quad\quad\ \   +(K\sin\varphi\cos\varphi+\Gamma'\cos 2\varphi)\cos\theta,\nonumber\\
\mbox{{\tt cd}{\rm{X}}}:\quad&\widetilde{J}_{ij}^{xz}=(\Gamma\sin\varphi+\Gamma'\cos\varphi)\cos2\theta\nonumber\\
&\quad\quad\ \   -(K\cos^2\varphi+\Gamma'\sin 2\varphi) \sin\theta\cos\theta,\nonumber\\
\label{eq:kh_vertex}
   \,      &\widetilde{J}_{ij}^{yz}=(\Gamma\cos\varphi-\Gamma'\cos\varphi)\sin\theta \\
   &\quad\quad\ \   -(K\sin\varphi\cos\varphi-\Gamma'\cos 2 \varphi)\cos\theta, \nonumber\\
\mbox{{\tt cd}{\rm{Y}}}:\quad&\widetilde{J}_{ij}^{xz}=(\Gamma\cos\varphi+\Gamma'\sin\varphi)\cos2\theta\nonumber\\
&\quad\quad\ \   -(K\sin^2\varphi+\Gamma'\sin 2\varphi)\sin\theta\cos\theta,\nonumber\\
   \,      &\widetilde{J}_{ij}^{yz}=-(\Gamma\sin\varphi-\Gamma'\cos\varphi)\sin\theta\nonumber\\
   &\quad\quad\ \   +(K\sin\varphi\cos\varphi+\Gamma'\cos 2 \varphi)\cos\theta,\nonumber\\
\mbox{{\tt ad}{\rm{Z}}({\tt bc}{\rm{Z}})}:\ &\widetilde{J}_{ij}^{xz}=
\big(\left(K-\Gamma\sin2\varphi\right)\sin\theta\cos\theta\nonumber\\
&\quad\quad\ \   +\Gamma' (\cos \varphi+\sin\varphi)\cos 2\theta\big)
\mbox{sign}(i-j),\nonumber\\
   \,      &\widetilde{J}_{ij}^{yz}=-\Gamma\cos2\varphi \cos\theta \nonumber\\
   &\quad\quad\ \     -\Gamma'(\cos\varphi-\sin \varphi)\sin \theta,\nonumber
\end{align}
where ${\tt a}$, ${\tt b}$, ${\tt c}$, ${\tt d}$ are the four sublattices of the zigzag state, and  X, Y, Z are the
three types of bonds, the bonds {\tt ab} and {\tt cd} can be of X and Y type, see Fig.~\ref{fig_strfac}.

The overall real-space coupling strength introduced in Eq.~(\ref{eq:3_scale}) can be written explicitly as 
\begin{equation}
\widetilde{V}_{\rm eff}=\frac{1}{6} \left(\big|\widetilde{V}_{{\tt ab}{\rm{X}}}\big|+
\big|\widetilde{V}_{{\tt ab}{\rm{Y}}}\big|+ \big|\widetilde{V}_{{\tt cd}{\rm{X}}}\big|+
\big|\widetilde{V}_{{\tt cd}{\rm{Y}}}\big|+2\big|\widetilde{V}_{{\tt ad}{\rm{Z}}}\big|\right).
\label{eq:3_scale_app}
\end{equation}

The out-of-plane tilt angle $\alpha$ of the zigzag structure relative to the basal plane of the honeycomb lattice
 is related to the polar and azimuthal angles as
\begin{equation}
\tan\alpha=\sqrt{2}\ \ \frac{\cos\varphi+\sin\varphi+\tan\theta}{\cos\varphi+\sin\varphi-2\tan\theta}.
\end{equation}
For all relevant values of $\Gamma$ considered in this work $\varphi\!=\!\pi/4$, see also Ref.~\cite{Chaloupka16}, 
so the spins are perpendicular to one of the bonds and 
\begin{equation}
\tan\alpha=\frac{\sqrt{2}+\tan\theta}{1-\sqrt{2}\tan\theta}.
\end{equation}
This also considerably simplifies the expressions in Eq.~(\ref{eq:kh_vertex}).

% ==============================================================================


\begin{thebibliography}{99}
\bibitem{F_footnote} There is a lower faint mode (F) below the single magnon mode (C) observed in ESR experiments 
\cite{Zvyagin17}. A comparison with the recent neutron scattering 
experiments \cite{Baltz19} suggests that this F-mode is a one-magnon signal at non-zero 
out-of-plane momentum $k_z=\pi$. The observation of such ``shifted'' modes is possible in spiral phases 
due to Dzyaloshinskii-Moriya interactions
\cite{ZvZh14}. In case of $\alpha$-RuCl$_3$ such terms are allowed in the  interplane couplings.
\bibitem{footnote3D} The interplane couplings discussed in Ref.~\cite{Vojta3D} can be
easily incorporated in our consideration as they do not enter the functional expression for $\Delta H_c$
and only modify our definition of $J_{03}$ in Eq.~(\ref{J03}). As such, the empirical value of 
$H_c^{(a)}$ will continue to restrict $J_{03}$, but not the exchanges individually. 
The interplane couplings may also play an additional role in stabilizing the zigzag phase.
\bibitem{footnote_C3} The anisotropy associated with the $C_3$ symmetry breaking can be 
parametrized by a reduction of all terms along the weaker bonds  by a factor $(1-\gamma)$. 
We have found that the critical fields depend linearly on $\gamma$, while the gap at the $M$ 
points reaches its experimental value at a rather small $\gamma\!\approx\!0.05$, corresponding to changes 
to the critical fields that are within the 10\% range. Since in our consideration the critical fields 
and other criteria are not fixed to a particular experimental value but are given a significant range,
we expect the effects of the $C_3$ symmetry breaking on our conclusions  to be minimal.
\end{thebibliography}
\end{document}